# Excellent and CO$_2$-resistant permeability of Ce$_{0.85}$Nd$_{0.1}$Cu$_{0.05}$O$_{2-\delta}$-Nd$_x$Sr$_{1-x}$Fe$_{1-y}$Cu$_y$O$_{3-\delta}$ dual-phase oxygen transport membranes

Chao Zhang [a, 1], Yue Zhu [b, 1], Xiaopeng Wang [a, 1], Yanhao Huang [a], Lingyong Zeng [a], Kuan Li [a], Peifeng Yu [a], Kangwang Wang [a], Longfu Li [a], Zaichen Xiang [a], Rui Chen [a], Xuefeng Zhu [b, *], Huixia Luo [a, *]

[a] School of Materials Science and Engineering, State Key Laboratory of Optoelectronic Materials and Technologies, Guangdong Provincial Key Laboratory of Magnetoelectric Physics and Devices, Key Lab of Polymer Composite & Functional Materials, Sun Yat-Sen University, Guangzhou 510275, China

[b] State Key Laboratory of Catalysis, Dalian Institute of Chemical Physics, Chinese Academy of Sciences, Dalian 116023, China

* Corresponding authors.

E-mail addresses: luohx7@mail.sysu.edu.cn (H. Luo), zhuxf@dicp.ac.cn (X. Zhu).

[1] These authors contributed equally to this work.

Keywords: Oxygen Separation, Oxygen Transport membranes, Dual-phase membranes, Carbon Dioxide Resistance

## Abstract

Oxygen transport membranes (OTMs) have provided great opportunities in the last decades but are suffering from the trade-off effect between stability and oxygen permeability. Here, we report a group of new planar dual-phase mixed ionic-electronic conducting (MIEC) OTMs consisting of Ce$_{0.85}$Nd$_{0.1}$Cu$_{0.05}$O$_{2-\delta}$ (CNCO) and Nd$_x$Sr$_{1-x}$Fe$_{1-y}$Cu$_y$O$_{3-\delta}$ (NSFCO; $x$ = 0.4, 0.6; $y$ = 0.05, 0.1) phases, showing excellent oxygen permeability while comparable CO$_2$-resistant stability. The substitution of Cu as a bifunctional additive decreases the sintering temperature and enhances bulk diffusion and oxygen permeability with the co-doping of Nd. The oxygen permeation fluxes reached 2.62 and 1.52 mL min$^{-1}$ cm$^{-2}$ at 1000 ºC through the optimal 60wt%Ce$_{0.85}$Nd$_{0.1}$Cu$_{0.05}$O$_{2-\delta}$-40wt%Nd$_{0.4}$Sr$_{0.6}$Fe$_{0.9}$Cu$_{0.1}$O$_{3-\delta}$ (CNCO-NSFCO41) composition with He and CO$_2$ sweeping, respectively, higher than all reported dense dual-phase OTMs. Such excellent CO$_2$-tolerant permeability meets the needs of potential industrial applications. Analysis with Zhu's oxygen permeation model shows lower bulk diffusion resistance of CNCO-NSFCO$_{41}$ than that of reported 60wt%Ce$_{0.85}$Pr$_{0.1}$Cu$_{0.05}$O$_{2-\delta}$-40wt%Pr$_{0.4}$Sr$_{0.6}$Fe$_{0.9}$Cu$_{0.1}$O$_{3-\delta}$ (CPCO-PSFCO$_{41}$) and more limitation by the interfacial exchange at high temperature. All the prepared OTMs also show good long-term stability over 100 hours in both atmospheres. Our results confirm the excellent oxygen permeability and stability under a high-concentration CO$_2$ atmosphere, providing a material candidate for CO$_2$ capture in oxyfuel combustion.

**Graphic Abstract**

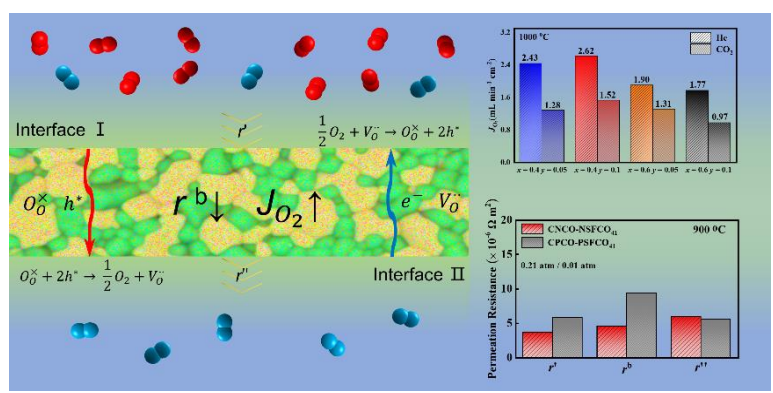

## 1. Introduction

Mixed ionic-electronic conducting (MIEC) oxygen transport membranes (OTMs), including single-phase membranes and dual-phase membranes, have received much attention in gas separation, solid oxide fuel cell (SOFC), and oxyfuel combustion based on CO$_2$ capture, and so on. Among single-phase perovskite OTMs, some Co-containing materials could show high oxygen permeability, such as Ba$_{1-x}$Sr$_x$Co$_y$Fe$_{1-y}$O$_{3-\delta}$ (BSCFO) and La$_{1-x}$Sr$_x$Co$_y$Fe$_{1-y}$O$_{3-\delta}$ (LSCFO) [1,2]. However, due to the facile reduction and easy evaporation of cobalt elements, Co-based single-phase perovskite OTMs have poor stability. Subsequently, cobalt-free perovskite OTMs are valued for their excellent stability. For instance, Ba$_{0.5}$Sr$_{0.5}$Zn$_{0.2}$Fe$_{0.8}$O$_{3-\delta}$ showed good long-term stability but much worse oxygen permeability.

Recent studies found that reducing the proportion of Sr by doping the A-site with lanthanides (Ln) can improve the structural stability, and selecting low-valence elements (such as Cu, Ni, etc.) at the B-site to replace Fe can improve the electrical conductivity of the material [3]. For instance, $Pr_{0.5}Sr_{0.5}Fe_{0.8}Cu_{0.2}O_{3-\delta}$ and $Nd_{0.5}Sr_{0.5}Fe_{0.8}Cu_{0.2}O_{3-\delta}$ [4,5] have been reported that their electrical conductivity can be compared with many cobalt-containing materials while their stability is much better. The 1.4-mm-thickness single-phase $Pr_{0.5}Sr_{0.5}Fe_{0.8}Cu_{0.2}O_{3-\delta}$ OTM exhibited an oxygen permeation flux ($J_{O_2}$) of 0.4 mL min$^{-1}$ cm$^{-2}$ at 900 °C under air/he gradient [6], higher than that of $Ba_{0.5}Sr_{0.5}Zn_{0.2}Fe_{0.8}O_{3-\delta}$ and other Co-free OTMs. These membranes often show Sr and Cu aggregation on surfaces after oxygen performance testing [3-8], causing the weakening of metal-oxygen (MO) bonds on membrane surfaces and forming more oxygen vacancies. However, aggregated Sr may form $SrCO_3$ purity in the $CO_2$ atmosphere, hindering oxygen ion transport and destroying the structural stability of OTMs.

Another option is to construct dual-phase MIEC OTMs, which can consist of fluorite oxides and perovskite oxides/other oxides/metals [9]. The former mainly acts as an oxygen ion conductor, while the latter is usually an electron conductor or a mixed ion-electron conductor. The two are evenly distributed and form a connected conductive network, which can improve charge exchange. $CO_2$-resistant fluorite oxide serves as the material's framework to effectively maintain a stable structure and oxygen ion pathways, significantly when carbonate disrupts oxygen ion transport in the perovskite phase. So, dual-phase OTMs usually have higher stability in $CO_2$. There are many dual-phase OTMs (e.g., $Ce_{0.9}Pr_{0.1}O_{2-\delta}$-$Pr_{0.6}Sr_{0.4}FeO_{3-\delta}$ (CPO-PSFO), $Ce_{0.9}Nd_{0.1}O_{2-\delta}$-$Nd_{0.6}Sr_{0.4}FeO_{3-\delta}$ (CNO-NSFO), and $Ce_{0.9}Pr_{0.1}O_{2-\delta}$-$Nd_xSr_{1-x}Fe_{0.9}Cu_{0.1}O_{3-\delta}$ (CPO-NSFCO)) have been designed and reported [10-12]. Nevertheless, they usually have poor oxygen permeability, which is lower than the value of industrial requirement (1 mL min$^{-1}$ cm$^{-2}$). Later, Fang et al. proposed a new MIEC-MIEC composite OTM (75 wt%$Ce_{0.85}Gd_{0.1}Cu_{0.05}O_{2-\delta}$-25 wt%$La_{0.6}Ca_{0.4}FeO_{3-\delta}$), which could improve the electronic conductivity and oxygen permeability by tiny Cu substitution into the fluorite phase [13]. Recently, our group discovered that the 60wt%$Ce_{0.85}Pr_{0.1}Cu_{0.05}O_{2-\delta}$-40wt%$Pr_{0.4}Sr_{0.6}Fe_{0.95}Cu_{0.05}O_{3-\delta}$ (CPCO-PSFCO) MIEC-MIEC OTM can achieve high fluxes of 1.60 and 0.98 mL min$^{-1}$ cm$^{-2}$ at 1000 °C under air/He or air/$CO_2$ gradient, respectively [14]. This shows that constructing the MIEC-MIEC dual-phase OTM can effectively improve oxygen permeability.

Herein, we further design and prepare a series of a novel 60wt%$Ce_{0.85}Nd_{0.1}Cu_{0.05}O_{2-\delta}$-40wt%$Nd_xSr_{1-x}Fe_{1-y}Cu_yO_{3-\delta}$ (CNCO-NSFCO; $x = 0.4, 0.6$; $y = 0.05, 0.1$) dual-phase MIEC OTMs based on the following aspects. (i) Nd is generally in +2 and +3 valence states, lower than the common valence states of Pr. Nd-containing materials among Ln-doped perovskite materials show high electrical conductivity at high temperatures [2,15]. Based on the common ion effects, the same element (Nd) in the two phases could also reduce the element diffusion between them [11]. (ii) The combined effect of Sr and Cu can also promote the forming of oxygen vacancies and facilitate the sintering of OTMs. (iii) $Ce_{0.9}Nd_{0.1}O_{2-\delta}$ also has high electrical conductivity among $Ce_{1-x}Ln_xO_{2-\delta}$ [16], and Cu doping can effectively improve its electronic conductivity. Higher conductivity could reduce the resistance of bulk diffusion in oxygen permeation. Therefore, it is expected that CNCO-NSFCO shows high oxygen permeability while having excellent stability by combining the advantages of the two components, CNCO and NSFCO. Our oxygen performance test demonstrates that the CNCO-NSFCO$_{41}$ ($x = 0.4$; $y = 0.1$) possesses super oxygen permeability with 2.62 mL min$^{-1}$ cm$^{-2}$ at 1000 °C under air/He gradient, which is higher than our previously reported 60wt%$Ce_{0.85}Pr_{0.1}Cu_{0.05}O_{2-\delta}$-40wt%$Pr_{0.4}Sr_{0.6}Fe_{0.9}Cu_{0.1}O_{3-\delta}$ (CPCO-PSFCO$_{41}$). Tests with regulated oxygen partial pressure and analysis with Zhu's oxygen permeation model show that one reason is the lower bulk diffusion resistance of CNCO-NSFCO$_{41}$ than that of reported CPCO-PSFCO$_{41}$. More importantly, when the sweep gas was switched to $CO_2$, the oxygen permeation flux was maintained at 1.52 mL min$^{-1}$ cm$^{-2}$, which was higher than other reported dense dual-phase OTMs and many stable single-phase OTMs. This work provides a material candidate for $CO_2$ capture in oxy-fuel combustion.

## 2. Experimental Procedures
### 2.1. Preparation of CNCO-NSFCO powders and membranes

The precursor powders of dual-phase CNCO-NSFCO ($x = 0.4, 0.6$; $y = 0.05, 0.1$) were prepared by a one-pot sol-gel method, and the detailed procedure has been reported in the previous literature [10]. The mass ratio of the two phases in this work was kept at 60 wt% and 40 wt%, which is obtained according to the previous work [17,18]. Nitrate hydrates corresponding to all metal cations were weighed and dissolved in distilled water, and the weighed mass was determined by multiplying the stoichiometric number by the molar ratio calculated by the mass ratio of two phases. The solution was stirred well while heating, after which citric acid monohydrate and ethylene glycol were added as chelating agents and surfactants, respectively. The molar ratio of total metal cations, citric acid monohydrate, and ethylene glycol in the solution was maintained at 1:2:2. After the water evaporates, the solution turns into a gel. A dry gel was obtained after drying at 140 °C for 24 hours, then ground and calcined at 600 °C for 8 hours to remove organics. The calcined powders were ground again and calcined at 950 °C for 12 hours to obtain the desired dual-phase CNCO-NSFCO precursor powders.

The precursor powders were uniaxially pressed under a pressure of 200 MPa for three minutes to obtain the green disks, using a circular stainless-steel mold (15 mm in diameter). Dense dual-phase MIEC OTMs were prepared after sintering at 1225 °C for 5 hours (with a heating and cooling rate of 1 °C min$^{-1}$).

## 2.2. Characterization of materials

Room temperature powder X-ray diffraction (PXRD, Rigaku MiniFlex 600, Cu Kα, 10 - 100° of 2θ range and 0.5°/min of scan rate) was used to characterize the crystal structure of the fresh dual-phase powders and OTM materials after testing. The diffraction patterns were refined by the FullProf Suite software with the Rietveld model. The phase structure diagrams were drawn by VESTA software [19], based on the refined data. Scanning Electron Microscopy (SEM, COXEM EM-30AX) was used to characterize the surface morphology and grain growth of the fresh and tested membranes. We also investigated the relative density and porosity of the sintered membranes using the Archimedes method. Finally, the element distribution of dual-phase grains and element content on the surface of the dual-phase membranes was characterized by Backscattered Scanning Electron Microscopy (BSEM) as well as Energy Dispersive X-ray Spectroscopy (EDXS).

## 2.3. Test of oxygen permeability and stability

The pre-obtained CNCO-NSFCO OTMs were used for oxygen permeability and stability tests with different sweep gases using a self-built high-temperature test device [20,21]. All the CNCO-NSFCO OTMs were polished to 0.6 mm, which will be sealed at one end of a long corundum tube with high-temperature ceramic glue (Huitian, Hubei, China) or silver. The outer side of the corundum tube is the feed side, and the inner side is the sweep side. The effective working area of OTMs is approximately 0.95 mm$^2$, calculated from the diameter of the inner wall of tubes (D = 11 mm). Although the application of high-temperature ceramic glue would bring a little error, we believe that the effective working area is the same during each test, and it has been confirmed with the vernier caliper after tests. During the test, 150 mL min$^{-1}$ of synthetic air (79 vol% $N_2$ and 21 vol% $O_2$) was used as feed gas to keep the oxygen partial pressure at 0.21. 1 mL min$^{-1}$ Ne (99.999 vol% Ne) + 49 mL min$^{-1}$ He (99.999 vol% He)/$CO_2$ (99.999 vol% $CO_2$) was used as the sweep gas to promote the surface exchange of oxygen, according to the requirements of tests. Ne in the sweep gas was used as the internal standard gas, and the flow rate and concentration of Ne assisted the total flow rate of the exhaust gas. The flow rates of all gases were adjusted by mass flow controllers (MFC, Sevenstar, Beijing,

China), and the exhaust gas obtained from all reactions was analyzed by gas chromatography (GC, Agilent 7890B, USA) for their compositions and concentrations of various gases. Exhaust gas typically includes permeated oxygen, leaked oxygen, nitrogen, and sweep gas. Due to the inconvenience of sealing at high temperatures, the leakage of a small amount of air in the oxygen permeability test of disk-shaped OTMs is a common problem. The leaked oxygen concentration should be less than 10% of the total oxygen concentration in the exhaust gas and be removed from the oxygen permeation flux ($J_{O_2}$) by the following equation (**Equation 1**) [22]:

$$J_{O_2} = \left(C_{O_2} - \frac{0.21}{0.79} \times C_{N_2}\right) \times \frac{F}{S} \tag{1}$$

where $J_{O_2}$ is the calculated oxygen permeation flux through the OTM, while $C_{O_2}$ and $C_{N_2}$ represent the concentrations of oxygen and nitrogen in the exhaust gas obtained by GC, respectively. $F$ represents the overall flow rate of the exhaust gas, and $S$ is the effective working area of each OTM mentioned above.

## 2.4. Analysis with Zhu's oxygen permeation model

To further study the oxygen permeability of the Nd-containing $Ce_{0.85}Nd_{0.1}Cu_{0.05}O_{2-\delta}$-$Nd_{0.4}Sr_{0.6}Fe_{0.9}Cu_{0.1}O_{3-\delta}$ (CNCO-NSFCO41; $x = 0.4$, $y = 0.1$) dual-phase MIEC OTMs, we compared it with the Pr-containing $Ce_{0.85}Pr_{0.1}Cu_{0.05}O_{2-\delta}$-$Pr_{0.4}Sr_{0.6}Fe_{0.9}Cu_{0.1}O_{3-\delta}$ (CPCO-PSFCO41). Zhu's oxygen permeation model was used to analyze the oxygen permeation of some dual-phase OTMs successfully [23]. The model is based on the following three assumptions. The detailed formula derivation has been reported [24]: (1) The transport properties of oxygen ions and electrons/holes do not vary with position or oxygen chemical potential in a given region. (2) The oxygen concentration polarization resistance on gas-solid interfaces of both sides is entirely negligible. (3) All reaction steps, including adsorption, dissociation, charge transfer, surface diffusion and incorporation of oxygen ions into the lattice are carried out under isothermal conditions. Basic fitting equations are as follows:

$$r^{tot} = r' + r^b + r'' = -\frac{RT\ln\frac{P_l}{P_h}}{16F^2 J_{O_2}} \tag{2}$$

$$r' = r_0'(P_h/P_0)^{-0.5} \tag{3}$$

$$r'' = r_0''(P_l/P_0)^{-0.5} \tag{4}$$

Where $r^{tot}$, $r^b$, $r'$ and $r''$ are the total permeation resistance, bulk diffusion resistance, and interfacial exchange resistances on the feed side and the sweep side, respectively. $R$, $T$, $F$ and $J_{O_2}$ are the gas constant, temperature, Faraday constant, and oxygen permeation flux, respectively. $P_l$ and $P_h$ are the lower and higher $P_{O_2}$ of two sides of the membranes, respectively. $P_0$, $r_0'$ and $r_0''$ represent $P_{O_2}$ of 1 atm, and permeation resistance constants at $P_{O_2}$ of 1 atm on the feed side and the sweep side, respectively.

Silver rings (99.99%) were used to seal OTMs here to minimize the leakage of oxygen, while the test temperatures were 900, 850, and 800 °C. At each test temperature, five different $P_h$ and four different $P_l$ were obtained by adjusting the ratio of pure $O_2$ in the feed gas and flow rates of the sweep gas, respectively. Then 20 groups of data were fit to the permeation model, and we could get $r^b$, $r'$ and $r''$ at a certain temperature. In this part of the test, the feed gas was changed to a mixture of $N_2$ (99.999 vol% $N_2$) and $O_2$ (99.999 vol% $O_2$) in five different proportions, which was controlled by MFC to regulate $P_h$ on the feed side. At the same time, $P_l$ on the sweep side was regulated by changing the He flow rate.

## 3. Results and Discussions
### 3.1. Characterizations of materials

XRD was used to investigate the crystal structure of our fresh CNCO-NSFCO powders obtained by calcination at 950 °C for 10 hours. As shown in **Figure 1**, all powders are mainly composed of fluorite CNCO and perovskite NSFCO. The crystal structure diagrams of two main phases (CNCO phase and NSFCO phase) were drawn with VESTA software (**Figure S1**). A trace amount of CuO impurity can be observed in all materials, resulting in two faint diffraction peaks at 35 ° and 39 °, which do not belong to the two main phases. The space groups of the CNCO phase and NSFCO phase are Fm-3m (Cubic) and Pbnm (Orthorhombic), respectively, which were clarified by XRD refinement. Among all the doped components, the crystal symmetry of CNCO and NSFCO did not change, which can be obtained from the refined lattice parameters of the two phases in **Table S1**.

The surface morphologies, structures, and element distribution of fresh CNCO–NSFCO membranes were characterized by BSEM (**Figure 2a**), XRD (**Figure 2b**) and SEM-EDXS (**Figure S2**). The XRD patterns of membranes of all components were still mainly composed of diffraction peaks corresponding to the cubic fluorite phase and the orthorhombic perovskite phase (**Figure 2b**). Comparing the XRD patterns of fresh powders (**Figure 1**), there was no significantly enhanced diffraction peak of CuO or diffraction peak of impurities in that of sintered membranes. This showed that there was no significant segregation of CuO or other impurities during the sintering process. No through-holes or pores appear on the surface of our fresh membranes, and all grain boundaries are visible. Compared with the Cu-free CPO-PSFO (sintered at 1400 °C [11]), our membranes were sintered at a lower temperature, which is beneficial to reduce the overall production cost of OTMs. The results of the Archimedes method also prove that the relative density is about 90% (**Table S2**), and the sintered compactness is good. Moreover, both phases show larger grains in our membranes than Cu-free membranes, which is related to the introduction of Cu as a sintering aid. Larger crystal grains could improve the bulk diffusion of oxygen ions in our materials, thereby enhancing oxygen permeability.

In the BSEM images (**Figure 2a**), the colors of the two phases are distinguished, showing their complementary structures. The color of the CNCO phase region is brighter, while the color of the NSFCO phase region is darker. This is because the average atomic mass of the elements in the CNCO phase is larger, resulting in more backscattered electrons, which is positively related to the atomic mass. The regions of the fluorite phase form a continuous network structure, which is beneficial to the stability of the dual-phase membrane and the smooth conduction of oxygen ions. This also shows that the mass ratio of the two phases we choose is appropriate [18].

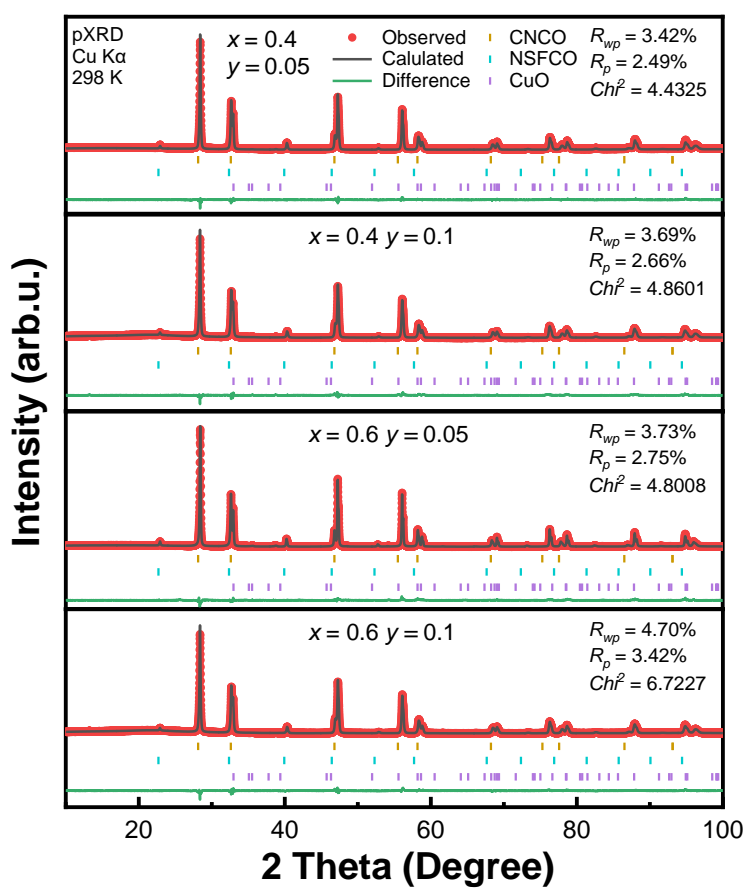

**Figure 1.** XRD refinements of fresh powders obtained by calcination at 950 °C for 10 hours. (CNCO-NSFCO; $x = 0.4, 0.6$; $y = 0.05, 0.1$)

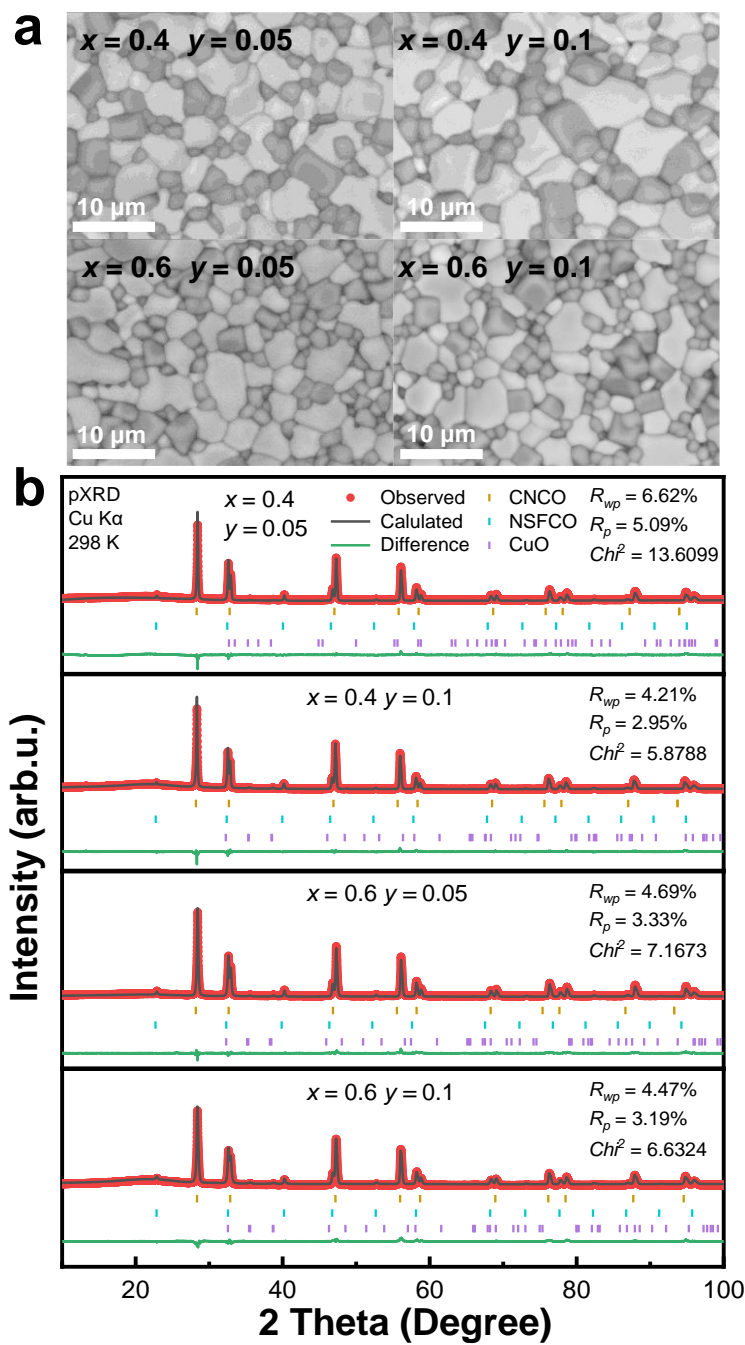

**Figure 2.** (a) BSEM images, (b) XRD refinements of membranes sintered at 1225 °C for 5 hours. (CNCO-NSFCO; x = 0.4, 0.6; y = 0.05, 0.1)

EDXS images (**Figure S2**) further suggest that elements in the CNCO phase and the NSFCO phase are distributed complementary, and the unique elements (Ce or Fe and Sr) in the two phases have no obvious interdiffusion. The measured and theoretical values of the atomic ratio of each element also show no significant difference in fresh membranes (**Table S3**).

Unlike the powder XRD results, no apparent aggregation of Cu element was found on the surface of membranes at this time.

The above results suggest that a series of dense dual-phase CNCO-NSFCO OTM materials have been successfully prepared.

**3.2. Oxygen permeability test**

The 0.6 mm-thickness dense membranes were used to test the oxygen permeability under He and $CO_2$ atmospheres. **Figures 3a** and **3b** are the $J_{O_2}$-T diagrams through dual-phase CNCO-NSFCO with He and $CO_2$ sweeping, respectively, presenting the high oxygen permeability of our OTM materials. The $J_{O_2}$ of all dual-phase OTMs increases with temperature regardless of whether the sweep gas was used. This is because the temperature could promote the two main processes in the oxygen permeation reaction, including surface exchange and bulk diffusion [25,26]. Among them, the optimal compound of $Ce_{0.85}Nd_{0.1}Cu_{0.05}O_{2-\delta}$-$Nd_{0.4}Sr_{0.6}Fe_{0.9}Cu_{0.1}O_{3-\delta}$ (CNCO-NSFCO41; $x = 0.4$, $y = 0.1$) exhibited the highest oxygen permeability, reaching the $J_{O_2}$ of 2.62 mL min$^{-1}$ cm$^{-2}$ and 1.52 mL min$^{-1}$ cm$^{-2}$ at 1000 °C under air/He gradient and air/$CO_2$ gradient, respectively. In addition, the oxygen permeability of all membranes under He sweeping is stronger than that under $CO_2$ sweeping. Since impurities such as strontium carbonate ($SrCO_3$) reported in previous work did not appear in our materials (**Figure S7**) [14,27], this could be attributed to the stronger adsorption of $CO_2$ than that of He on the membrane surface, which hinders the oxygen exchange [28]. The apparent activation energies ($E_a$) are also shown in **Figures 3a** and **3b**, obtained from the Arrhenius plots of oxygen permeability (**Figure S3**). All dual-phase membranes have higher apparent activation energies ($E_a$) under air/$CO_2$ gradient, which is consistent with the lower $J_{O_2}$ at this time. The CNCO-NSFCO41 membrane has the lowest apparent $E_a$ under the air/He gradient, which is also consistent with its optimal oxygen permeability.

In addition, it is worth noting that except for the $J_{O_2}$ of the CNCO-NSFCO61 membrane in $CO_2$ atmosphere is 0.97 mL min$^{-1}$ cm$^{-2}$, all the other three components of OTMs can reach a $J_{O_2}$ more than 1 mL min$^{-1}$ cm$^{-2}$ at 1000 °C with $CO_2$ sweeping, much better than the Cu-free CNO-NSFO (0.26 mL min$^{-1}$ cm$^{-2}$ or 0.21 mL min$^{-1}$ cm$^{-2}$ with He/$CO_2$ sweeping in 950 °C) [11]. It is mainly attributed to substituting Cu and regulating Nd/Sr content. Adding Cu greatly enhanced the electronic conductivity of the CNO phase to construct MIEC-MIEC OTMs. This avoids the hindrance of oxygen exchange by the lack of electronic conductivity of the fluorite phase and extends TPB to the entire material surface [13]. In the perovskite phase, since Cu ions have lower valence states and smaller Cu-O bond energy, Cu substitution reduces the average bond energy (ABE) between B-site ions and O ions, resulting in more oxygen vacancies [14,29].

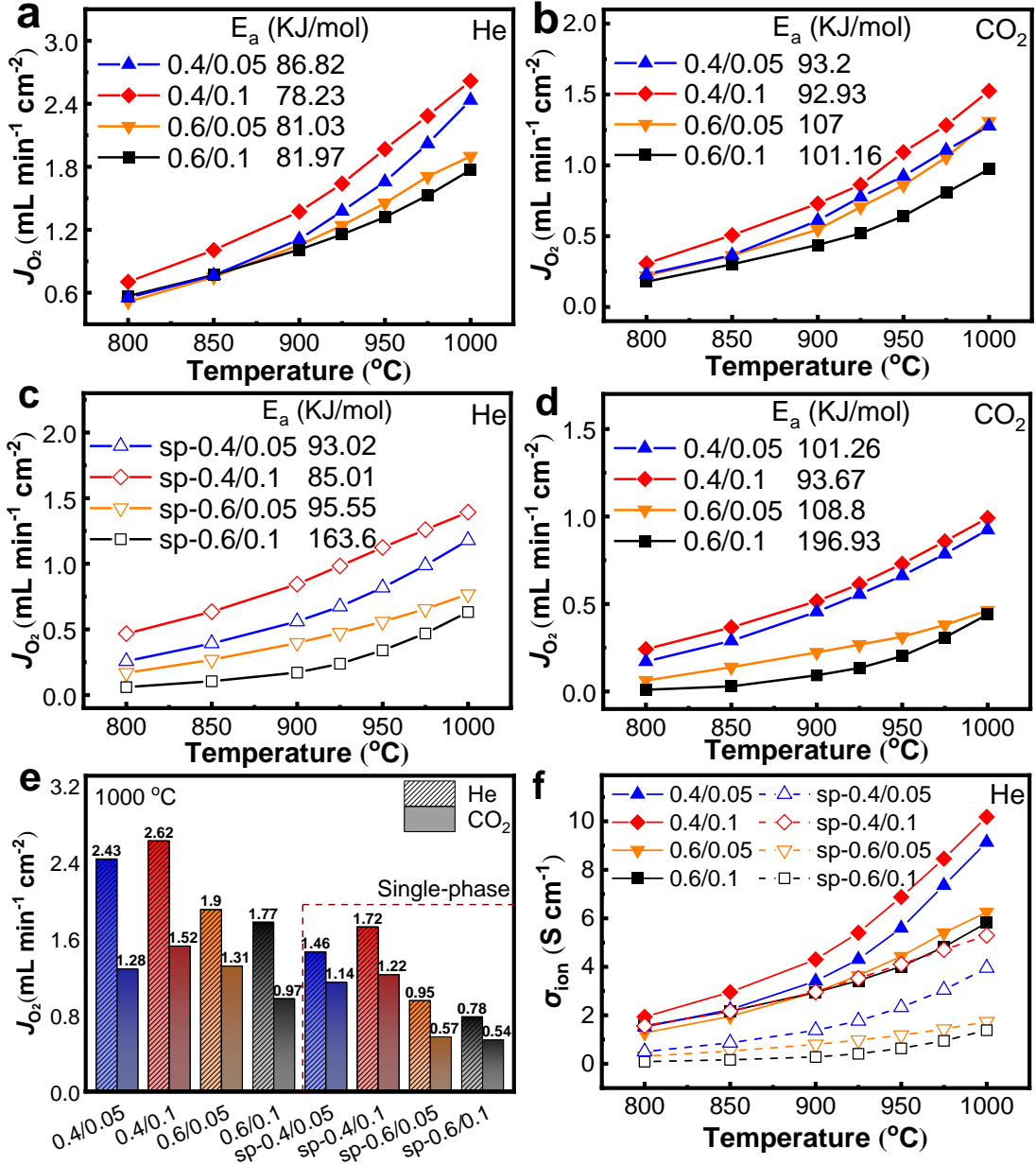

**Figure 3.** (a, b) $J_{O_2}$-T diagrams of dual-phase OTMs; (c, d) $J_{O_2}$-T diagrams of single-phase perovskite OTMs. (e) $J_{O_2}$ of all membranes at 1000 °C with two sweeping gases; (f) Temperature-dependence ionic conductivity calculated by **Equation 7**. (Marked as *x/y* for dual phase 60 wt%$Ce_{0.85}Nd_{0.1}Cu_{0.05}O_{2-\delta}$-40 wt%$Nd_xSr_{1-x}Fe_{1-y}Cu_yO_{3-\delta}$ ($x$ = 0.4, 0.6; $y$ = 0.05, 0.1), and sp-*x/y* for single-phase $Nd_xSr_{1-x}Fe_{1-y}Cu_yO_{3-\delta}$ ($x$ = 0.4, 0.6; $y$ = 0.05, 0.1).)

Due to the excellent oxygen permeability of our materials, which is rare in dual-phase MIEC OTMs, we investigated the oxygen permeability of membranes with four single-phase components of $Nd_xSr_{1-x}Fe_{1-y}Cu_yO_{3-\delta}$ (NSFCO; $x$ = 0.4, 0.6; $y$ = 0.05, 0.1) for comparison. The same sintering temperature (1225 °C) as the dual-phase membranes were chosen to obtain membranes with relative densities above 95%. And the $J_{O_2}$ of all single-phase membranes is lower than that of dual-phase membranes with the corresponding contents (**Figures 3c** and **3d**), which may be due to the addition of Cu in the fluorite phase to construct the MIEC-MIEC dual-phase membranes. The three-phase boundaries (TPBs), namely boundaries among the electronic conducting phase, the ionic conducting phase, and the gaseous phase, are extended over the entire membrane surface. This enables smoother charge exchange between phases, which significantly improves oxygen permeability [13,30,31]. In addition, the oxygen permeability of the component with an Nd content of 0.4 in the single-phase OTMs is also stronger than that of OTMs with an Nd content of 0.6, which is the same as that in dual-phase OTMs. **Figure S4** shows the XRD patterns of our single-phase membranes (compared with CNCO-NSFCO$_{41}$), and the peak positions are the same as those in the dual-phase materials. SEM-EDXS shows no holes at all on the surface (**Figure S5**), and the crystal grains on the membrane surface of the two components with 10 at% Cu are significantly larger than those of the components with 5 at% Cu, which shows the excellent effect of Cu as a sintering aid. But this is not obvious enough in the dual-phase membranes, probably because of the small mass ratio and compression strain of the perovskite phase in them [18].

For a better comparison of the oxygen permeability of OTMs with different compositions, the $J_{O_2}$ through all membranes at 1000 °C with two different sweep gases are shown in **Figure 3e**. In both atmospheres, the OTM of CNCO-NSFCO$_{41}$ components showed the highest oxygen permeability, followed by the CNCO-NSFCO$_{45}$ OTM, indicating that the oxygen permeability of OTMs with the Nd content of 0.4 was stronger than that of OTMs with Nd content of 0.6. This is because of the important effect of chemical doping on the crystal symmetry and oxygen vacancies of the perovskite phase (ABO$_3$ structure). The ionic radius of $Sr^{2+}$ (0.144 nm) is larger than that of $Nd^{3+}$ (0.127 nm), so the increasing Sr content at the A site is beneficial to enhancing unit cell symmetry, oxygen vacancy generation, and electrical conductivity [3,32]. In addition, Sr may partially compete with the precipitation of Cu in the material [3,5], and there will be more Cu accumulation in the material with less Sr content. This may make more precipitation of CuO in the material with an Nd content of 0.6 and affect the structure of the material, especially in CNCO-NSFCO$_{61}$ with more Cu content, resulting in weaker oxygen permeability than CNCO-NSFCO$_{65}$.

When bulk diffusion is the controlling step in the oxygen permeation process, $J_{O_2}$ can be calculated by the Wagner equation (**Equation 6**):

$$J_{O_2} = \frac{RT}{16F^2L} \int_{P_l}^{P_h} \frac{\sigma_{ion}\sigma_e}{\sigma_{ion}+\sigma_e} d(\ln P_{O_2}) \qquad (6)$$

Where *R, T, F,* and *L* represent a gas constant, temperature, Faraday constant, and membrane thickness, respectively. $P_{O_2}$ represents the partial pressure of oxygen. $P_l$ and $P_h$ are the lower and higher $P_{O_2}$ of two sides of the membranes, respectively. $\sigma_{ion}$ is the ionic conductivity and $\sigma_e$ is the electronic conductivity. For MIEC OTMs, $\sigma_e$ is often much greater than $\sigma_{ion}$, so Equation 6 could be simplified to **Equation 7**:

$$J_{O_2} = \frac{RT}{16F^2L} \sigma_{ion} \ln\frac{P_h}{P_l} \qquad (7)$$

The ionic conductivity of the material can be obtained from the measured data with **Equation 7**. **Figure 3f** shows the ionic conductivity versus temperature for all CNCO-NSFCO and NSFCO OTMs under He sweeping. Consistent with the trend of oxygen flux, the ionic conductivity increases with increasing temperature. The optimal CNCO-NSFCO$_{41}$ reaches the ionic conductivity of 10.18 S cm$^{-1}$ at 1000 °C, consistent with its highest oxygen permeability. However, this may exceed the actual value because the assumption that bulk diffusion is the control step ignores the resistance of interfacial exchange. Our OTMs may be more limited by the interfacial exchange process at high temperatures, which will be discussed in the next section.

**Figure 4** shows $J_{O_2}$ through various reported high-performance planar OTM materials [12-14,20,28,33-44]. Although there are some differences in the test conditions, we can see that CNCO-NSFCO$_{41}$ performs the best oxygen permeability among reported dual-phase OTMs, including Co-containing Ce$_{0.9}$Gd$_{0.1}$O$_{2-\delta}$–Ba$_{0.5}$Sr$_{0.5}$Co$_{0.8}$Fe$_{0.2}$O$_{3-\delta}$, Ce$_{0.9}$Pr$_{0.1}$O$_{2-\delta}$–Pr$_{0.6}$Sr$_{0.4}$Fe$_{0.5}$Co$_{0.5}$O$_{3-\delta}$ and Cu-containing Ce$_{0.8}$Sm$_{0.2}$O$_{2-\delta}$–Sm$_{0.3}$Sr$_{0.7}$Fe$_{0.8}$Cu$_{0.2}$O$_{3-\delta}$ [33,35,41]. With the same test temperature of 1000 °C and thickness of 0.6 mm, CNCO-NSFCO$_{41}$ performs better than single-phase Pr$_{0.6}$Sr$_{0.4}$Co$_{0.5}$Fe$_{0.5}$O$_{3-\delta}$ whatever the sweep gas is. BSCFO possesses the optimal oxygen permeability among reported OTMs, but CNCO-NSFCO$_{41}$ has advantages over BSCFO and doped Ba$_{0.5}$Sr$_{0.5}$Co$_{0.78}$Fe$_{0.2}$W$_{0.02}$O$_{3-\delta}$ when CO$_2$ sweeping. The permeability is also comparable to that of other doped stable single-phase OTMs [28,38-40], especially in the CO$_2$ atmosphere. This CO$_2$-resistant oxygen permeability is valuable in potential industrial applications such as oxy-fuel combustion.

Interestingly, this is better than the CPCO-PSFCO$_{41}$ we recently developed as well. Comparing the XRD refinement of their powders shows that the perovskite phase in CNCO-NSFCO$_{41}$ has larger unit cell parameters and volume (**Table S4**), which is conducive to the generation and transport of oxygen vacancies. The tilting angle (Φ) of the $BO_6$ octahedron in CNCO-NSFCO41 is a little smaller, showing higher spatial symmetry [45]. Higher symmetry structure is more conducive to the movement of oxygen ions at high temperatures because of more possible moving directions. In addition, the absolute value of the formation enthalpy of $Nd_2O_3$ is smaller than that of $Pr_2O_3$ [46], which indicates that the bond between $Nd^{3+}$ and $O^{2-}$ is weaker. Weaker metal-oxygen bonds are also conducive to the generation of oxygen vacancies and the improvement of oxygen permeability [47].

It should be noted that the measurement on our dual-phase OTMs was based on planar membranes without any coating or reductive sweeping gases, like CO and $H_2$. Our materials could likely be improved by preparing hollow fiber membranes or asymmetric membranes (reducing thickness) or by coating porous layers (increasing surface area) [17,28,31,48].

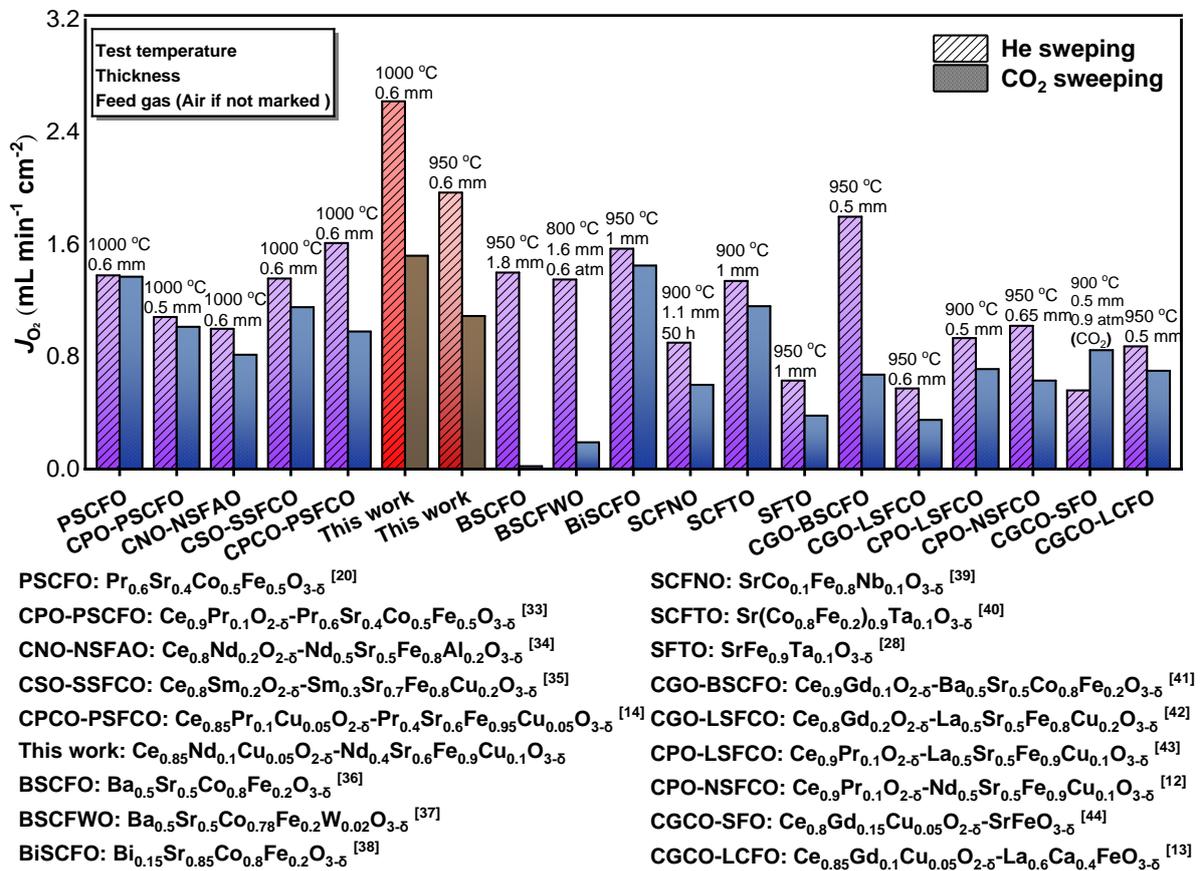

**Figure 4.** $J_{O_2}$ through various reported high-performance planar OTM materials.

## 3.3. Analysis with Zhu's oxygen permeation model

To further study the oxygen permeation processes, we compared the resistance of interfacial exchange and bulk diffusion of CNCO-NSFCO41 and CPCO-PSFCO41 by Zhu's oxygen permeation model (**Figure 5a**, **5b**, and **5c**). Zhu's oxygen permeation model was used to analyze the oxygen permeation of some dual-phase OTMs successfully and it was stated in the supporting information [23]. The permeation resistances decrease with the increasing temperature because all the sub-steps of oxygen permeation are thermally activated. Among them, bulk diffusion resistance ($r^b$) decreased rapidly with increasing temperature, which exhibits the strongest thermal activation properties. Especially the $r^b$ of CNCO-NSFCO$_{41}$ decreases to be lower than the interfacial exchange resistance on the sweep side ($r''$) at 900 ℃, which indicates that the oxygen permeation process of CNCO-NSFCO$_{41}$ at high temperatures may be more limited by the surface exchange process. Comparing the permeation resistances of the two materials, $r''$ is close at different temperatures, which indicates that the two materials perform similarly in OER applications. However, both $r^b$ and the interfacial exchange resistance on the feed side ($r'$) of CNCO-NSFCO$_{41}$ are significantly lower than those of CPCO-PSFCO$_{41}$. The lower $r^b$ is related to the higher electrical conductivity, because, among the A-site lanthanide-doped perovskite materials, Nd-doped materials usually exhibit higher electrical conductivity at high temperatures than Pr-doped materials [2,4,5,15]. Nd-doped CeO$_2$ also performs better conductivity at high temperatures [16], and high electrical conductivity is beneficial for bulk diffusion according to the Wagner equation (**Equation 6**).

**Figure 5d**, **5e**, and **5f** show the $J_{O_2}$ under different oxygen partial pressures on two sides. At all temperatures, higher $P_h$ and lower $P_l$ lead to greater $J_{O_2}$. With the increasing temperature, the oxygen permeability of CNCO-NSFCO$_{41}$ at the same oxygen partial pressure gradually exceeds that of CPCO-PSFCO$_{41}$, which is attributed to the above-mentioned higher conductivity of Nd-containing materials at high temperatures. At the same time, with higher temperatures, the increase of $J_{O_2}$ caused by reducing $P_l$ is more obvious. This is consistent with the analytic result that $r^b$ is less than $r''$ at high temperatures, which proves the greater control effect of interfacial exchange when the temperature is higher than 900 °C. Lower $r'$ is beneficial to the adsorption and reduction of oxygen on the feed side, which suggests that Nd-containing materials may perform better in the ORR reaction. Relatively, Nd$_{0.4}$Sr$_{0.6}$Co$_{0.8}$Fe$_{0.2}$O$_{3-\delta}$ exhibited the best catalytic activity for oxygen reduction among Ln$_{0.4}$Sr$_{0.6}$Co$_{0.8}$Fe$_{0.2}$O$_{3-\delta}$ (Ln = La, Pr, Nd, Sm, Gd) materials used as the cathode of solid oxide fuel cells (SOFCs) [2].

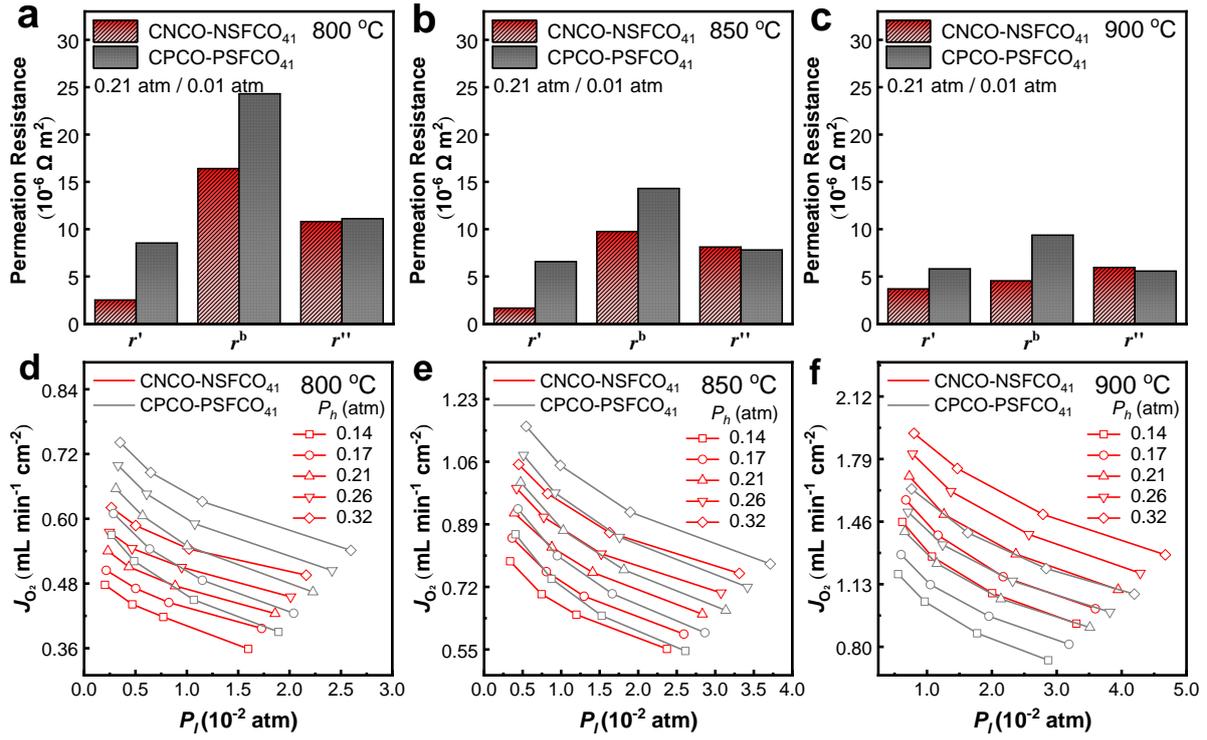

**Figure 5.** Permeation resistances (Conditions: 0.21 atm and 0.01 atm of oxygen partial pressure on two sides) of dual-phase 60wt%$Ce_{0.85}Nd_{0.1}Cu_{0.05}O_{2-\delta}$-40wt%$Nd_{0.4}Sr_{0.6}Fe_{0.9}Cu_{0.1}O_{3-\delta}$ (CNCO-NSFCO$_{41}$) and 60wt%$Ce_{0.85}Pr_{0.1}Cu_{0.05}O_{2-\delta}$-40wt%$Pr_{0.4}Sr_{0.6}Fe_{0.9}Cu_{0.1}O_{3-\delta}$ (CPCO-PSFCO$_{41}$) at (a) 800 °C; (b) 850 °C; (c) 900 °C; Corresponding $J_{O_2}$ under different oxygen partial pressures on two sides at (d) 800 °C; (e) 850 °C; (f) 900 °C.

### 3.4. Stability test

To investigate the stability of CNCO-NSFCO materials in a high-temperature hypoxic environment (pure Ar) and corrosive environment (pure $CO_2$), we heat-treated fresh dual-phase powders obtained by calcination at 950 °C in both atmospheres. Three temperatures of 800 °C, 900 °C, and 1000 °C were selected for each atmosphere, and room temperature XRD was used to characterize the change in the crystal structure of materials after heat treatment. **Figure S6** shows the XRD patterns of samples after 24 hours of treatment in pure Ar. The structures of the two main phases in the samples at the three temperatures did not change. The previously reported phenomenon of CuO impurity conversion into $Cu_2O$ [13,14] was not found, in favor of the stability of the material in a high-temperature and low-oxygen environment. **Figure S7** shows the XRD patterns of powders treated in pure $CO_2$ for 24 hours. The phase structure did not change compared to the fresh powders. In particular, the occurrence of $SrCO_3$ impurities reported in other literature [14,49] was not found at all three temperatures. Carbonate formation marks the chemical reaction between OTMs and $CO_2$, rather than just adsorption, which hinders oxygen transport inside the OTMs and destabilizes the membrane structure. Our materials had no carbonate formation, demonstrating their good $CO_2$ stability, especially compared to the CPCO-PSFCO material. According to Lewis acid-base theory, the relative acidity of $Nd_2O_3$ is stronger than that of $Pr_2O_3$ and SrO. The increase in overall relative acidity makes carbonate formation more difficult, thereby enhancing the $CO_2$ resistance of the materials [50].

Further, we continued the stability test for 100 hours after the oxygen permeability test of OTMs under both atmospheres. **Figure 6** shows $J_{O_2}$ versus time through the dual-phase CNCO-NSFCO membranes. When He was used as the sweep gas, the $J_{O_2}$ through the CNCO-NSFCO61 decreased from 1.70 mL min$^{-1}$ cm$^{-2}$ to 1.60 mL min$^{-1}$ cm$^{-2}$ in 100 hours, which may be related to the reduction of CuO, while the oxygen permeability of other membranes decreased slightly. When the sweep gas was $CO_2$, the oxygen permeability of all membranes decreased slowly with time. After 100 hours of testing, the $J_{O_2}$ of CNCO-NSFCO41 decreased from 1.52 mL min$^{-1}$ cm$^{-2}$ to 1.46 mL min$^{-1}$ cm$^{-2}$, while the $J_{O_2}$ of CNCO-NSFCO45 decreased from 1.30 mL min$^{-1}$ cm$^{-2}$ to 1.27 mL min$^{-1}$ cm$^{-2}$. This shows that our OTMs have good $CO_2$ resistance, which needs further research and improvement.

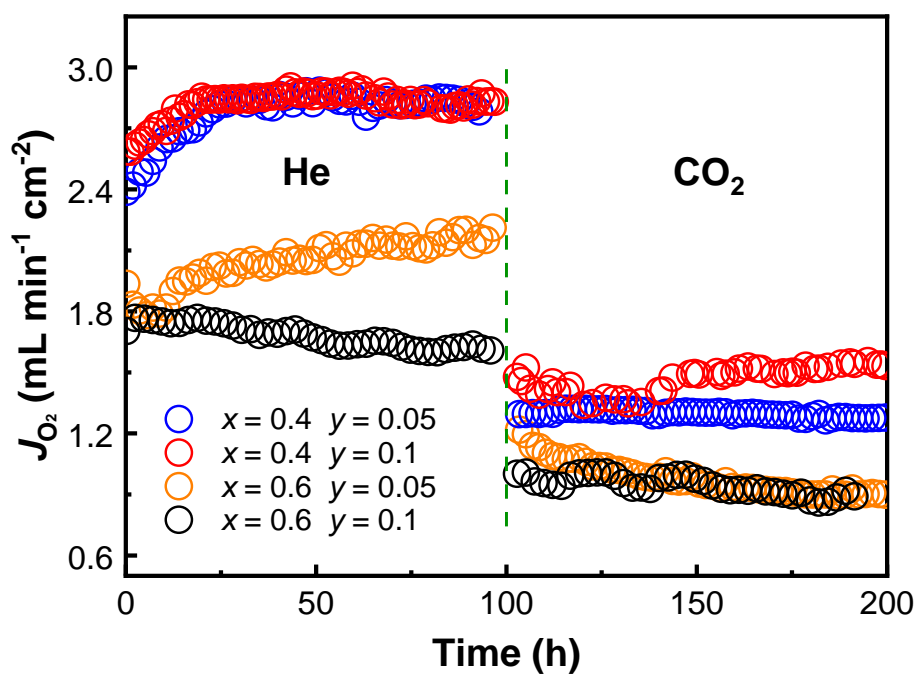

**Figure 6.** $J_{O_2}$ versus time through dual-phase 60 wt%$Ce_{0.85}Nd_{0.1}Cu_{0.05}O_{2-\delta}$-40 wt%$Nd_xSr_{1-x}Fe_{1-y}Cu_yO_{3-\delta}$ (CNCO-NSFCO; $x$ = 0.4, 0.6; $y$ = 0.05, 0.1) OTMs at 1000 °C with two different sweep gases, sealed with high-temperature ceramic glue.

The composition of the two sides of spent membranes often changes because some elements aggregate and precipitate on the surface of membranes during the long-term reaction [51,52]. Therefore, we used SEM-EDXS to characterize the surface topography and elemental composition of both sides of membranes after the oxygen permeation test. **Figure S8** shows the surface topography of both sides of spent membranes. The grain boundaries on the surface were blurred due to the grinding before testing. At the same time, some large grains are precipitated on the surface of the feed side, but there is no such phenomenon on the sweep side. Combined with the EDXS diagrams of the feed side (**Figure S9**), these grains can be judged as the aggregation of Cu. In addition, the XRD results of both sides of spent membranes show that the material is still mainly composed of fluorite CNCO and perovskite NSFCO phases (**Figure S10**). No carbonate formation was found, which is consistent with the heat treatment results for powders. The other diffraction peaks correspond to CuO impurities and high-temperature ceramic glue, indicating that the considerable grain precipitation in **Figure S8** and **Figure S9** may be CuO. **Figure S11** shows the EDXS mappings of the sweep side, and no large crystal grains were found like the feed side. **Table S5** shows the surface element contents of spent membranes. Although Cu aggregation also existed on the feed side in this work, there was no significant increase in Sr content on the sweep side compared with that in fresh membranes (**Table S3**). The dynamic generation and decomposition of carbonate in CPCO-PSFCO promoted the transport and segregation of cations to surfaces of the membranes [14,53], and resulted in an abnormally high proportion of Cu and Sr on the membrane surface after the tests. The above results indicate that similar phenomena may not exist in Nd-containing membranes.

## 4. Conclusion

In this work, a new group of Nd and Cu co-doped dual-phase mixed conductor (MIEC) oxygen transport membrane (OTM) materials with the chemical formula of 60 wt%$Ce_{0.85}Nd_{0.1}Cu_{0.05}O_{2-\delta}$-40 wt%$Nd_xSr_{1-x}Fe_{1-y}Cu_yO_{3-\delta}$ (CNCO-NSFCO; $x = 0.4, 0.6$; $y = 0.05, 0.1$) were obtained by the one-pot sol-gel method and had a low sintering temperature of 1225 ºC. The optimal $Ce_{0.85}Nd_{0.1}Cu_{0.05}O_{2-\delta}$-$Nd_{0.4}Sr_{0.6}Fe_{0.9}Cu_{0.1}O_{3-\delta}$ (CNCO-NSFCO41; $x = 0.4$, $y = 0.1$) membrane exhibited a high oxygen permeation flux ($J_{O_2}$) of 2.62 mL min$^{-1}$ cm$^{-2}$ under air/He gradient at 1000 ºC, which is rare among dual-phase OTMs and also better than the single-phase OTMs of corresponding components. When the sweep gas is $CO_2$, the $J_{O_2}$ also reaches 1.52 mL min$^{-1}$ cm$^{-2}$, which is higher than the fundamental requirements of industrial application. The membranes of all components have worked stably for over 100 hours under the two sweep gases, respectively, showing good stability. After 100 hours of $CO_2$ sweeping, the $J_{O_2}$ of CNCO-NSFCO41 only decreased by 0.06 mL min$^{-1}$ cm$^{-2}$, indicating it has good $CO_2$ resistance. Studies on single-phase components show that the oxygen permeability of CNCO-NSFCO is stronger than that of single-phase NSFCO at the same sintering temperature, and this excellent oxygen permeability is rare in dual-phase OTMs.

Compared with Cu-doped 60 wt%Ce$_{0.85}$Pr$_{0.1}$Cu$_{0.05}$O$_{2-\delta}$-40 wt%Pr$_{0.4}$Sr$_{0.6}$Fe$_{0.9}$Cu$_{0.1}$O$_{3-\delta}$ (CPCO-PSFCO41), analysis with Zhu's permeation model show that CNCO-NSFCO41 has lower bulk diffusion resistance and interfacial exchange resistance on the feed side, which is beneficial to the oxygen permeation process. And the oxygen permeation might be more limited by the interfacial exchange at high temperatures (≥ 900 °C). This finding provides a potential high-performance $CO_2$-resistant membrane candidate for oxygen separation and $CO_2$ capture in the combustion process.

**Acknowledgments**

This work is supported by the National Natural Science Foundation of China (12274471, 11922415), the Guangdong Basic and Applied Basic Research Foundation (2022A1515011168, 2019A1515011718). The experiments reported were conducted at the Guangdong Provincial Key Laboratory of Magnetoelectric Physics and Devices, No. 2022B1212010008.

**Reference**

[1] Z.P. Shao, S.M. Haile, A high-performance cathode for the next generation of solid-oxide fuel cells, Nature 431 (2004) 170-173. https://doi.org/10.1038/nature02863.
[2] H.Y. Tu, Y. Takeda, N. Imanishi, O. Yamamoto, Ln$_{0.4}$Sr$_{0.6}$Co$_{0.8}$Fe$_{0.2}$O$_{3-\delta}$ (Ln = La, Pr, Nd, Sm, Gd) for the electrode in solid oxide fuel cells, Solid State Ionics 117 (1999) 277-281. https://doi.org/10.1016/S0167-2738(98)00428-7.
[3] J. Lu, Y.M. Yin, J.W. Yin, J.C. Li, J. Zhao, Z.F. Ma, Role of Cu and Sr in Improving the Electrochemical Performance of Cobalt-Free Pr$_{1-x}$Sr$_x$Fe$_{1-y}$Cu$_y$O$_{3-\delta}$ Cathode for Intermediate Temperature Solid Oxide Fuel Cells, J. Electrochem. Soc. 163 (2016) F44. https://doi.org/10.1149/2.0181602jes.
[4] J. Lu, Y.M. Yin, Z.F. Ma, Preparation and characterization of new cobalt-free cathode Pr$_{0.5}$Sr$_{0.5}$Fe$_{0.8}$Cu$_{0.2}$O$_{3-\delta}$ for IT-SOFC, Int. J. Hydrog. Energy 38 (2013) 10527-10533. https://doi.org/10.1016/j.ijhydene.2013.05.164.
[5] J.W. Yin, Y.M. Yin, J. Lu, C.M. Zhang, N.Q. Minh, Z.F. Ma, Structure and Properties of Novel Cobalt-Free Oxides Nd$_x$Sr$_{1-x}$Fe$_{0.8}$Cu$_{0.2}$O$_{3-\delta}$ (0.3 ≤ $x$ ≤ 0.7) as Cathodes of Intermediate Temperature Solid Oxide Fuel Cells, J. Phys. Chem. C 118 (2014) 13357-13368. https://doi.org/10.1021/jp500371w.
[6] Z.T. Wang, W. Liu, Y.S. Wu, W.P. Sun, W. Liu, C.Q. Wang, A novel cobalt-free $CO_2$-stable perovskite-type oxygen permeable membrane, J. Membr. Sci. 573 (2019) 504-510. https://doi.org/10.1016/j.memsci.2018.12.014.
[7] M. Alifanti, J. Kirchnerova, B. Delmon, D. Klvana, Methane and propane combustion over lanthanum transition-metal perovskites: role of oxygen mobility, Appl. Catal., A 262 (2004) 167-176. https://doi.org/10.1016/j.apcata.2003.11.024.

[8] I. Kaus, H.U. Anderson, Electrical and thermal properties of $La_{0.2}Sr_{0.8}Cu_{0.1}Fe_{0.9}O_{3-\delta}$ and $La_{0.2}Sr_{0.8}Cu_{0.2}Fe_{0.8}O_{3-\delta}$, Solid State Ionics 129 (2000) 189-200. https://doi.org/10.1016/S0167-2738(99)00325-2.

[9] C. Zhang, J. Sunarso, S. Liu, Designing $CO_2$-resistant oxygen-selective mixed ionic–electronic conducting membranes: guidelines, recent advances, and forward directions, Chem. Soc. Rev. 46 (2017) 2941-3005. https://doi.org/10.1039/C6CS00841K.

[10] H.X. Luo, H.Q. Jiang, T. Klande, Z.W. Cao, F.Y. Liang, H.H. Wang, J.r. Caro, Novel cobalt-free, noble metal-free oxygen-permeable $40Pr_{0.6}Sr_{0.4}FeO_{3-\delta}$–$60Ce_{0.9}Pr_{0.1}O_{2-\delta}$ dual-phase membrane, Chem. Mater. 24 (2012) 2148-2154. https://doi.org/10.1021/cm300710p.

[11] H.X. Luo, T. Klande, Z.W. Cao, F.Y. Liang, H.H. Wang, J. Caro, A $CO_2$-stable reduction-tolerant Nd-containing dual phase membrane for oxyfuel $CO_2$ capture, J. Mater. Chem. A 2 (2014) 7780-7787. https://doi.org/10.1039/C3TA14870J.

[12] G.X. Chen, Z.J. Zhao, M. Widenmeyer, R.J. Yan, L. Wang, A. Feldhoff, A. Weidenkaff, Synthesis and Characterization of 40 wt % $Ce_{0.9}Pr_{0.1}O_{2-\delta}$–60 wt % $Nd_xSr_{1-x}Fe_{0.9}Cu_{0.1}O_{3-\delta}$ Dual-Phase Membranes for Efficient Oxygen Separation, Membranes 10 (2020) 183. https://doi.org/10.3390/membranes10080183.

[13] W. Fang, F.Y. Liang, Z.W. Cao, F. Steinbach, A. Feldhoff, A Mixed Ionic and Electronic Conducting Dual-Phase Membrane with High Oxygen Permeability, Angew. Chem. Int. Ed. 54 (2015) 4847-4850. https://doi.org/10.1002/anie.201411963.

[14] X.P. Wang, Y.H. Huang, D.C. Li, L.Y. Zeng, Y.Y. He, M. Boubeche, H.X. Luo, High oxygen permeation flux of cobalt-free Cu-based ceramic dual-phase membranes, J. Membr. Sci. 633 (2021) 119403. https://doi.org/10.1016/j.memsci.2021.119403.

[15] Y.H. Chen, Y.J. Wei, H.H. Zhong, J.F. Gao, X.Q. Liu, G.Y. Meng, Synthesis and electrical properties of $Ln_{0.6}Ca_{0.4}FeO_{3-\delta}$ (Ln = Pr, Nd, Sm) as cathode materials for IT-SOFC, Ceram. Int. 33 (2007) 1237-1241. https://doi.org/10.1016/j.ceramint.2006.03.035.

[16] M. Balaguer, C. Solís, J.M. Serra, Structural–Transport Properties Relationships on $Ce_{1-x}Ln_xO_{2-\delta}$ System (Ln = Gd, La, Tb, Pr, Eu, Er, Yb, Nd) and Effect of Cobalt Addition, J. Phys. Chem. C 116 (2012) 7975-7982. https://doi.org/10.1021/jp211594d.

[17] L. Shi, S. Wang, T.N. Lu, Y. He, D. Yan, Q. Lan, Z.A. Xie, H.Q. Wang, M.R. Li, J. Caro, H.X. Luo, High $CO_2$-tolerance oxygen permeation dual-phase membranes $Ce_{0.9}Pr_{0.1}O_{2-\delta}$-$Pr_{0.6}Sr_{0.4}Fe_{0.8}Al_{0.2}O_{3-\delta}$, J. Alloys Compd. 806 (2019) 500-509. https://doi.org/10.1016/j.jallcom.2019.07.281.

[18] J.Y. Wang, Q.K. Jiang, D.D. Liu, L.M. Zhang, L.L. Cai, Y. Zhu, Z.W. Cao, W.P. Li, X.F. Zhu, W.S. Yang, Effect of inner strain on the performance of dual-phase oxygen permeable membranes, J. Membr. Sci. 644 (2022) 120142. https://doi.org/10.1016/j.memsci.2021.120142.

[19] K. Momma, F. Izumi, VESTA 3 for three-dimensional visualization of crystal, volumetric and morphology data, J. Appl. Crystallogr. 44 (2011) 1272-1276. https://doi.org/10.1107/S0021889811038970.

[20] K. Partovi, F.Y. Liang, O. Ravkina, J. Caro, High-Flux Oxygen-Transporting Membrane $Pr_{0.6}Sr_{0.4}Co_{0.5}Fe_{0.5}O_{3-\delta}$: $CO_2$ Stability and Microstructure, ACS Appl. Mater. Interfaces 6 (2014) 10274-10282. https://doi.org/10.1021/am501657j.

[21] H.X. Luo, K. Efimov, F.Y. Liang, H.H. Wang, J. Caro, $CO_2$-tolerant oxygen-permeable $Fe_2O_3$-$Ce_{0.9}Gd_{0.1}O_{2-\delta}$ dual phase membranes, Ind. Eng. Chem. Res. 50 (2011) 13508–13517. https://doi.org/10.1021/ie200517t.

[22] Y.H. Huang, C. Zhang, X.P. Wang, D.C. Li, L.Y. Zeng, Y.Y. He, P.F. Yu, H.X. Luo, High $CO_2$ resistance of indium-doped cobalt-free 60 wt.%$Ce_{0.9}Pr_{0.1}O_{2-\delta}$-40 wt.%$Pr_{0.6}Sr_{0.4}Fe_{1-x}In_xO_{3-\delta}$ oxygen transport membranes, Ceram. Int. 48 (2022) 415-426. https://doi.org/10.1016/j.ceramint.2021.09.117.

[23] Y. Zhu, W.P. Li, Y. Liu, X.F. Zhu, W.S. Yang, Selection of oxygen permeation models for different mixed ionic-electronic conducting membranes, AIChE J. 63 (2017) 4043-4053. https://doi.org/10.1002/aic.15718.

[24] X.F. Zhu, H.Y. Liu, Y. Cong, W.S. Yang, Permeation model and experimental investigation of mixed conducting membranes, AlChE J. 58 (2012) 1744-1754. https://doi.org/10.1002/aic.12710.

[25] S. Kim, Y.L. Yang, A.J. Jacobson, B. Abeles, Diffusion and surface exchange coefficients in mixed ionic electronic conducting oxides from the pressure dependence of oxygen permeation, Solid State Ionics 106 (1998) 189-195. https://doi.org/10.1016/S0167-2738(97)00492-X.

[26] C. Wagner, Equations for transport in solid oxides and sulfides of transition metals, Prog. Solid State Chem. 10 (1975) 3-16. https://doi.org/10.1016/0079-6786(75)90002-3.

[27] O. Ravkina, T. Klande, A. Feldhoff, Investigation of carbonates in oxygen-transporting membrane ceramics, J. Membr. Sci. 480 (2015) 31-38. https://doi.org/10.1016/j.memsci.2015.01.042.

[28] J.W. Zhu, S.B. Guo, Z.Y. Chu, W.Q. Jin, $CO_2$-tolerant oxygen-permeable perovskite-type membranes with high permeability, J. Mater. Chem. A 3 (2015) 22564-22573. https://doi.org/10.1039/C5TA04598C.

[29] Z.L. Liu, K. Li, H.L. Zhao, K. Świerczek, High-performance oxygen permeation membranes: Cobalt-free $Ba_{0.975}La_{0.025}Fe_{1-x}Cu_xO_{3-\delta}$ ceramics, J. Materiomics 5 (2019) 264-272. https://doi.org/10.1016/j.jmat.2019.01.013.

[30] X.P. Wang, L. Shi, Y.H. Huang, L.Y. Zeng, M. Boubeche, D.C. Li, H.X. Luo, $CO_2$-Tolerant Oxygen Permeation Membranes Containing Transition Metals as Sintering Aids with High Oxygen Permeability, Processes 9 (2021) 528. https://doi.org/10.3390/pr9030528.

[31] E. Magnone, J. Chae, J.H. Park, Synthesis and oxygen permeation properties of $Ce_{0.8}Sm_{0.2}O_{2-d}$ - $Sm_{1-x}Sr_xCu_{0.2}Fe_{0.8}O_{3-d}$ dual-phase ceramic membranes: effect of Strontium contents and Pd coating layer, Ceram. Int. 44 (2018) 12948-12956. https://doi.org/10.1016/j.ceramint.2018.04.110.

[32] Y.T. Liu, X.Y. Tan, K. Li, Mixed Conducting Ceramics for Catalytic Membrane Processing, Cat. Rev. - Sci. Eng. 48 (2006) 145-198. https://doi.org/10.1080/01614940600631348.

[33] F.Y. Liang, H.X. Luo, K. Partovi, O. Ravkina, Z.W. Cao, Y. Liu, J. Caro, A novel $CO_2$-stable dual phase membrane with high oxygen permeability, Chem. Commun. 50 (2014) 2451-2454. https://doi.org/10.1039/C3CC47962E.

[34] K. Partovi, M. Bittner, J. Caro, Novel $CO_2$-tolerant Al-containing membranes for high-temperature oxygen separation, J. Mater. Chem. A 3 (2015) 24008-24015. https://doi.org/10.1039/C5TA04405G.

[35] K. Partovi, C.H. Rüscher, F. Steinbach, J. Caro, Enhanced oxygen permeability of novel Cu-containing $CO_2$-tolerant dual-phase membranes, J. Membr. Sci. 503 (2016) 158-165. https://doi.org/10.1016/j.memsci.2016.01.019.

[36] Z.P. Shao, W.S. Yang, Y. Cong, H. Dong, J.H. Tong, G.X. Xiong, Investigation of the permeation behavior and stability of a $Ba_{0.5}Sr_{0.5}Co_{0.8}Fe_{0.2}O_{3-\delta}$ oxygen membrane, J. Membr. Sci. 172 (2000) 177-188. https://doi.org/10.1016/S0376-7388(00)00337-9.

[37] M.P. Popov, I.A. Starkov, S.F. Bychkov, A.P. Nemudry, Improvement of $Ba_{0.5}Sr_{0.5}Co_{0.8}Fe_{0.2}O_{3-\delta}$ functional properties by partial substitution of cobalt with tungsten, J. Membr. Sci. 469 (2014) 88-94. https://doi.org/10.1016/j.memsci.2014.06.022.

[38] M. Li, H.J. Niu, J. Druce, H. Téllez, T. Ishihara, J.A. Kilner, H. Gasparyan, M.J. Pitcher, W. Xu, J.F. Shin, L.M. Daniels, L.A.H. Jones, V.R. Dhanak, D.Y. Hu, M. Zanella, J.B. Claridge, M.J. Rosseinsky, A $CO_2$-Tolerant Perovskite Oxide with High Oxide Ion and Electronic Conductivity, Adv. Mater. 32 (2020) 1905200. https://doi.org/10.1002/adma.201905200.

[39] Z.G. Wang, N. Dewangan, S. Das, M.H. Wai, S. Kawi, High oxygen permeable and $CO_2$-tolerant $SrCo_xFe_{0.9-x}Nb_{0.1}O_{3-\delta}$ ($x$ = 0.1–0.8) perovskite membranes: Behavior and mechanism, Sep. Purif. Technol. 201 (2018) 30-40. https://doi.org/10.1016/j.seppur.2018.02.046.

[40] W. Chen, C.S. Chen, L. Winnubst, Ta-doped $SrCo_{0.8}Fe_{0.2}O_{3-\delta}$ membranes: Phase stability and oxygen permeation in $CO_2$ atmosphere, Solid State Ionics 196 (2011) 30-33. https://doi.org/10.1016/j.ssi.2011.06.011.

[41] J. Xue, Q. Liao, Y.Y. Wei, Z. Li, H.H. Wang, A $CO_2$-tolerance oxygen permeable $60Ce_{0.9}Gd_{0.1}O_{2-\delta}$–$40Ba_{0.5}Sr_{0.5}Co_{0.8}Fe_{0.2}O_{3-\delta}$ dual phase membrane, J. Membr. Sci. 443 (2013) 124-130. https://doi.org/10.1016/j.memsci.2013.04.067.

[42] G.X. Chen, Z.J. Zhao, M. Widenmeyer, T. Frömling, T. Hellmann, R. Yan, F. Qu, G. Homm, J.P. Hofmann, A. Feldhoff, A. Weidenkaff, A comprehensive comparative study of $CO_2$-resistance and oxygen permeability of 60 wt % $Ce_{0.8}M_{0.2}O_{2-\delta}$ (M = La, Pr, Nd, Sm, Gd) - 40 wt % $La_{0.5}Sr_{0.5}Fe_{0.8}Cu_{0.2}O_{3-\delta}$ dual-phase membranes, J. Membr. Sci. 639 (2021) 119783. https://doi.org/10.1016/j.memsci.2021.119783.

[43] G.X. Chen, B.J. Tang, M. Widenmeyer, L. Wang, A. Feldhoff, A. Weidenkaff, Novel $CO_2$-tolerant dual-phase $Ce_{0.9}Pr_{0.1}O_{2-\delta}$ - $La_{0.5}Sr_{0.5}Fe_{0.9}Cu_{0.1}O_{3-\delta}$ membranes with high oxygen permeability, J. Membr. Sci. 595 (2020) 117530. https://doi.org/10.1016/j.memsci.2019.117530.

[44] W. Fang, F. Steinbach, C.S. Chen, A. Feldhoff, An Approach To Enhance the $CO_2$ Tolerance of Fluorite–Perovskite Dual-Phase Oxygen-Transporting Membrane, Chem. Mater. 27 (2015) 7820-7826. https://doi.org/10.1021/acs.chemmater.5b03823.

[45] S. Wang, L. Shi, M. Boubeche, H.Q. Wang, Z.A. Xie, W. Tan, Y. He, D. Yan, H.X. Luo, The effect of Fe/Co ratio on the structure and oxygen permeability of Ca-containing composite membranes, Inorg. Chem. Front. 6 (2019) 2885-2893. https://doi.org/10.1039/C9QI00822E.

[46] J.A. Dean, LANGE'S HANDBOOK OF CHEMISTRY, Mater. Manuf. Processes 5 (1990) 687-688. https://doi.org/10.1080/10426919008953291.

[47] K. Qiu, Y.L. Liu, J.K. Tan, T.L. Wang, G.R. Zhang, Z.K. Liu, W.Q. Jin, Fluorine-doped barium cobaltite perovskite membrane for oxygen separation and syngas production, Ceram. Int. 46 (2020) 27469-27475. https://doi.org/10.1016/j.ceramint.2020.07.235.

[48] Z.C. Zhang, K. Ning, Z. Xu, Q.K. Zheng, J.K. Tan, Z.K. Liu, Z.T. Wu, G.R. Zhang, W.Q. Jin, Highly efficient preparation of $Ce_{0.8}Sm_{0.2}O_{2-\delta}$–$SrCo_{0.9}Nb_{0.1}O_{3-\delta}$ dual-phase four-channel hollow fiber membrane via one-step thermal processing approach, J. Membr. Sci. 620 (2021) 118752. https://doi.org/10.1016/j.memsci.2020.118752.

[49] F. Miccio, A.N. Murri, E. Landi, High-Temperature Capture of $CO_2$ by Strontium Oxide Sorbents, Ind. Eng. Chem. Res. 55 (2016) 6696–6707. https://doi.org/10.1021/acs.iecr.6b00184.

[50] N.C. Jeong, J.S. Lee, E.L. Tae, Y.J. Lee, K.B. Yoon, Acidity Scale for Metal Oxides and Sanderson's Electronegativities of Lanthanide Elements, Angew. Chem. Int. Ed. 47 (2008) 10128-10132. https://doi.org/10.1002/anie.200803837.

[51] Y.M. Yin, M.W. Xiong, N.T. Yang, Z. Tong, Y.Q. Guo, Z.F. Ma, E. Sun, J. Yamanis, B.Y. Jing, Investigation on thermal, electrical, and electrochemical properties of scandium-doped $Pr_{0.6}Sr_{0.4}(Co_{0.2}Fe_{0.8})_{1-x}Sc_xO_{3-\delta}$ as cathode for IT-SOFC, Int. J. Hydrog. Energy 36 (2011) 3989-3996. https://doi.org/10.1016/j.ijhydene.2010.12.113.

[52] Y. Zhu, D.D. Liu, H.J. Jing, F. Zhang, X.B. Zhang, S.Q. Hu, L.M. Zhang, J.Y. Wang, L.X. Zhang, W.H. Zhang, B.J. Pang, P. Zhang, F.T. Fan, J.P. Xiao, W. Liu, X.F. Zhu, W.S. Yang, Oxygen activation on Ba-containing perovskite materials, Sci. Adv. 8 (2022) eabn4072. https://doi.org/10.1126/sciadv.abn4072.

[53] J.X. Yi, M. Schroeder, T. Weirich, J. Mayer, Behavior of $Ba(Co, Fe, Nb)O_{3-\delta}$ Perovskite in $CO_2$-Containing Atmospheres: Degradation Mechanism and Materials Design, Chem. Mater. 22 (2010) 6246-6253. https://doi.org/10.1021/cm101665r.

# Supplementary material

**Excellent and CO$_2$-resistant permeability of Ce$_{0.85}$Nd$_{0.1}$Cu$_{0.05}$O$_{2-\delta}$-Nd$_x$Sr$_{1-x}$Fe$_{1-y}$Cu$_y$O$_{3-\delta}$ dual-phase oxygen transport membranes**

Chao Zhang [a, 1], Yue Zhu [b, 1], Xiaopeng Wang [a, 1], Yanhao Huang [a], Lingyong Zeng [a], Kuan Li [a], Peifeng Yu [a], Kangwang Wang [a], Longfu Li [a], Zaichen Xiang [a], Rui Chen [a], Xuefeng Zhu [b, *], Huixia Luo [a, *]

[a] School of Materials Science and Engineering, State Key Laboratory of Optoelectronic Materials and Technologies, Guangdong Provincial Key Laboratory of Magnetoelectric Physics and Devices, Key Lab of Polymer Composite & Functional Materials, Sun Yat-Sen University, Guangzhou 510275, China

[b] State Key Laboratory of Catalysis, Dalian Institute of Chemical Physics, Chinese Academy of Sciences, Dalian 116023, China

* Corresponding authors.

E-mail addresses: luohx7@mail.sysu.edu.cn (H. Luo), zhuxf@dicp.ac.cn (X. Zhu).

[1] These authors contributed equally to this work.

**Table S1. Lattice parameters of the CNCO phase and NSFCO phase were obtained from the refinement of XRD results.**

|  | CNCO | $N_xS_{1-x}F_{1-y}C_yO$ | | |
|---|---|---|---|---|
|  | (*No.* 225:Fm-3m) | (*No.* 62:Pbnm) | | |
|  | a/Å | a/Å | b/Å | c/Å |
| $x = 0.4$ $y = 0.05$ | 5.43047 (12) | 5.4791(4) | 5.4642(6) | 7.7407(9) |
| $x = 0.4$ $y = 0.1$ | 5.42833 (11) | 5.4691(18) | 5.4712(16) | 7.7371(11) |
| $x = 0.6$ $y = 0.05$ | 5.43415 (7) | 5.4876(5) | 5.4809(4) | 7.7486(5) |
| $x = 0.6$ $y = 0.1$ | 5.43709 (10) | 5.48783(12) | 5.47483(13) | 7.74967(17) |

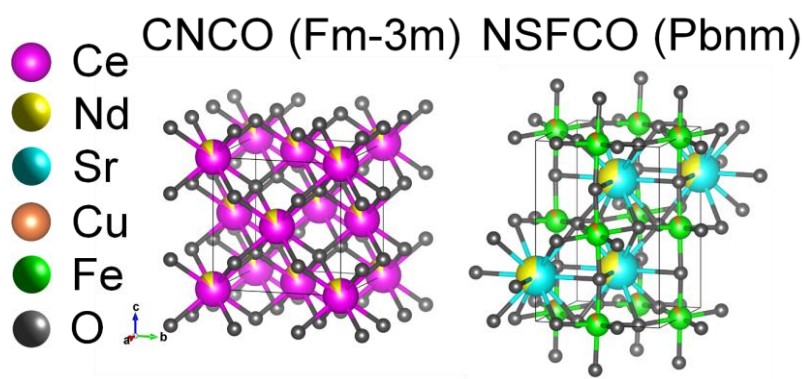

**Figure S1. The crystal structure diagrams of the CNCO phase and NSFCO phase were drawn with VESTA software.**

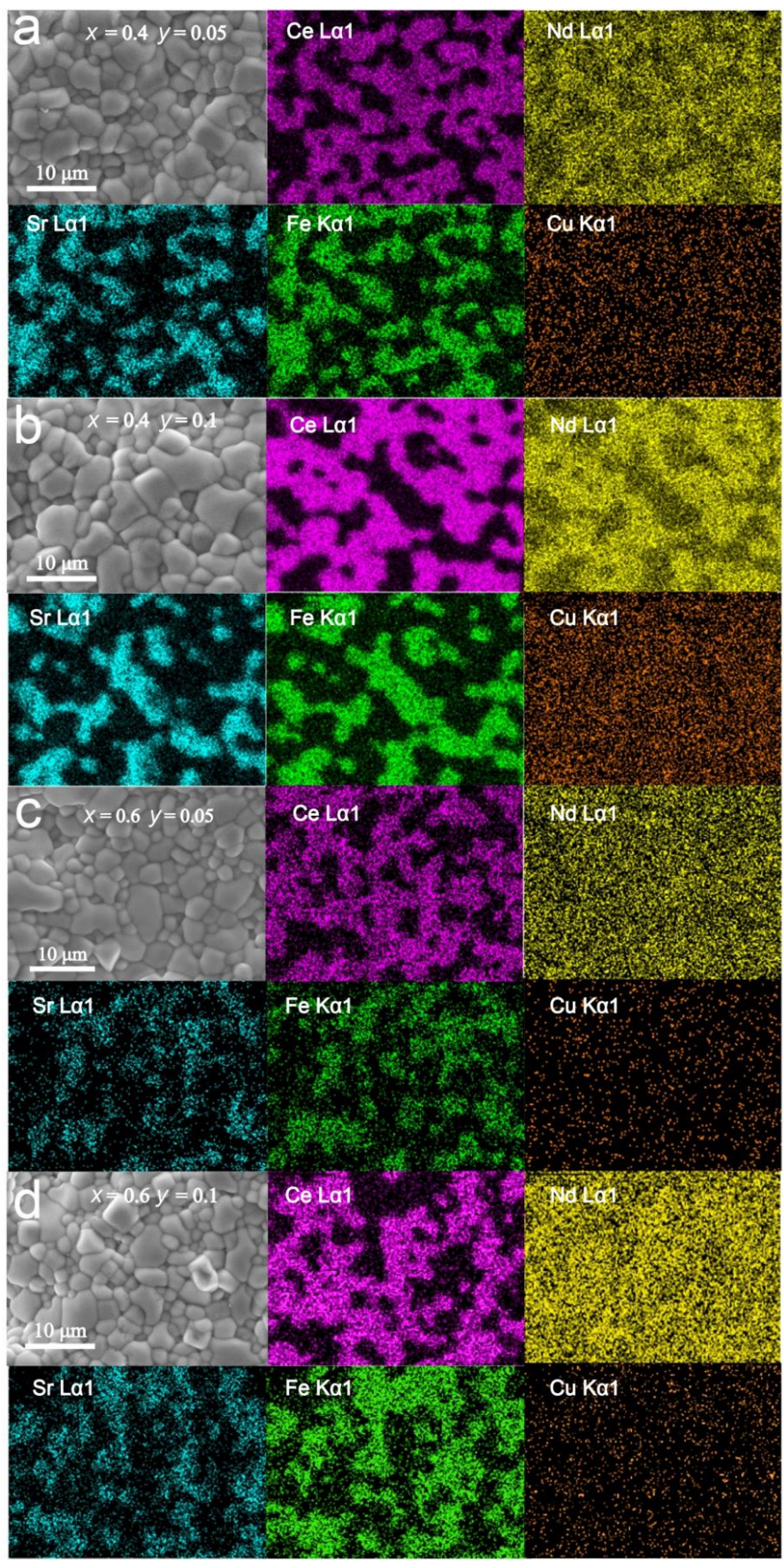

**Figure S2. Surface morphologies and element distribution on the surface of the sintered dual-phase CNCO-NSFCO ($x$ = 0.4, 0.6; $y$ = 0.05, 0.1) membranes.**

**Table S2. Relative density of fresh membranes of 60 wt%$Ce_{0.85}Nd_{0.1}Cu_{0.05}O_{2-\delta}$-40 wt%$Nd_xSr_{1-x}Fe_{1-y}Cu_yO_{3-\delta}$ (CNCO-NSFCO; $x$ = 0.4, 0.6; $y$ = 0.05, 0.1) sintered at 1225 ºC for 5 hours, obtained by Archimedes method.**

| Composition | Volume Density | Theoretical Density | Relative Density | Porosity |
|---|---|---|---|---|
| $x$ = 0.4 $y$ = 0.05 | 6.1236 | 6.6306 | 92.35% | 7.65% |
| $x$ = 0.4 $y$ = 0.1 | 5.9377 | 6.6460 | 89.34% | 10.66% |
| $x$ = 0.6 $y$ = 0.05 | 6.2370 | 6.7532 | 92.36% | 7.64% |
| $x$ = 0.6 $y$ = 0.1 | 6.1996 | 6.7544 | 91.79% | 8.21% |

**Table S3. Experimental and theoretical values of the atomic ratio of each element on the surface of fresh dual-phase membranes.**

| Composition | State | Element (Atomic percentage) | | | | |
|---|---|---|---|---|---|---|
| | | Ce | Nd | Fe | Sr | Cu |
| $x$ = 0.4 $y$ = 0.05 | Theoretical | 41.49% | 15.12% | 24.31% | 15.36% | 3.72% |
| | Experimental | 41.63% | 14.45% | 24.62% | 17.32% | 1.98% |
| $x$ = 0.4 $y$ = 0.1 | Theoretical | 41.53% | 15.11% | 23.01% | 15.34% | 5.00% |
| | Experimental | 42.28% | 15.02% | 23.23% | 16.38% | 3.19% |
| $x$ = 0.6 $y$ = 0.05 | Theoretical | 42.58% | 19.98% | 23.70% | 9.98% | 3.75% |
| | Experimental | 44.32% | 19.24% | 23.42% | 12.14% | 0.87% |
| $x$ = 0.6 $y$ = 0.1 | Theoretical | 42.62% | 19.97% | 22.44% | 9.97% | 5.00% |
| | Experimental | 43.88% | 19.43% | 22.18% | 11.87% | 2.64% |

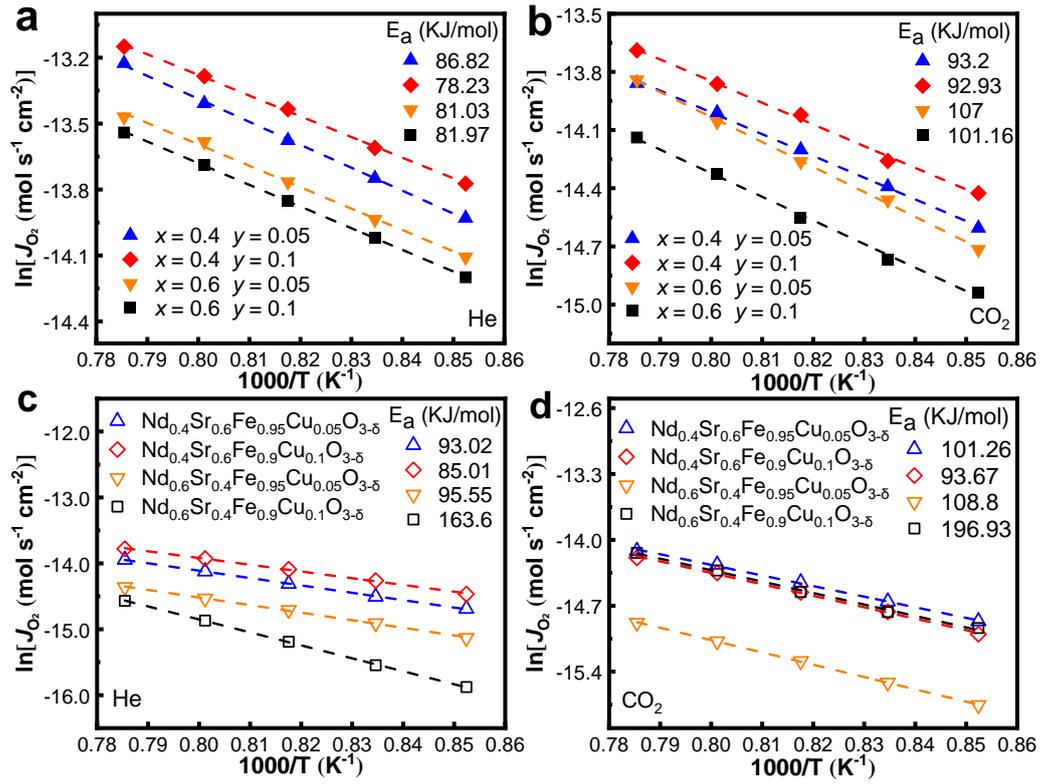

**Figure S3.** (a, b) $J_{O_2}$ through $Ce_{0.85}Nd_{0.1}Cu_{0.05}O_{2-\delta}$-$Nd_xSr_{1-x}Fe_{1-y}Cu_yO_{3-\delta}$ (NSFCO; $x$ = 0.4, 0.6; $y$ = 0.05, 0.1) dual-phase membranes and (c, d) the Arrhenius plots under (a, c) He, (b, d) $CO_2$ sweeping.

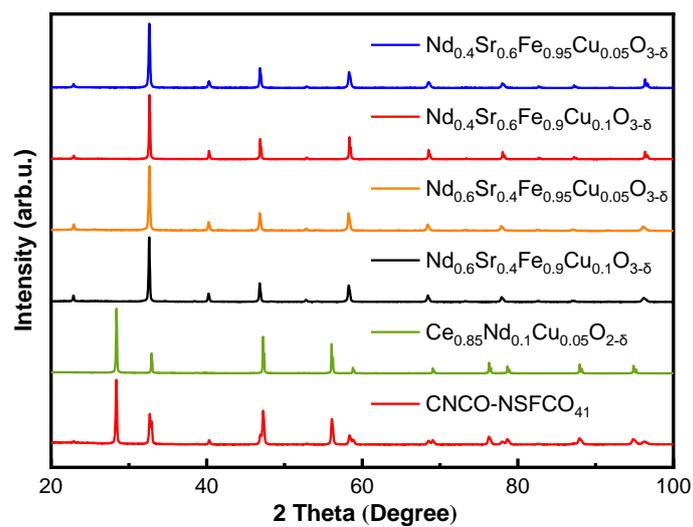

**Figure S4.** XRD patterns of single-phase and dual-phase OTM powders.

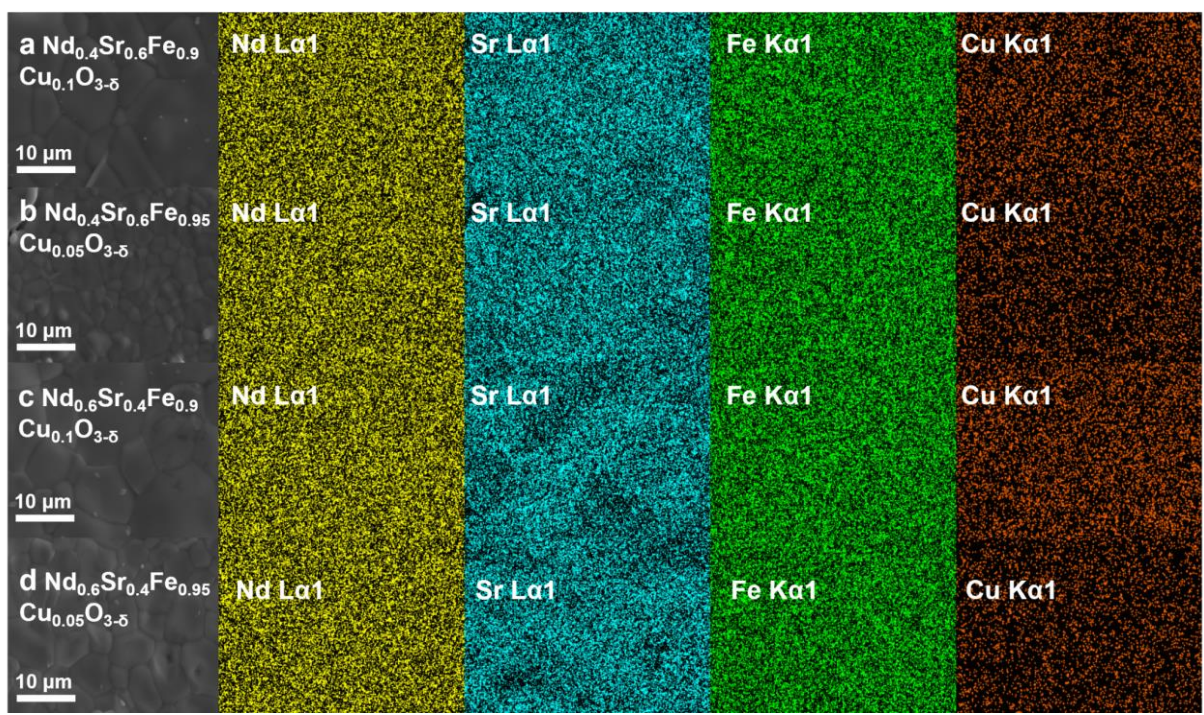

**Figure S5.** Surface morphology and element distribution of $Ce_{0.85}Nd_{0.1}Cu_{0.05}O_{2-\delta}$ and $Nd_xSr_{1-x}Fe_{1-y}Cu_yO_{3-\delta}$ (NSFCO; $x$ = 0.4, 0.6; $y$ = 0.05, 0.1) single-phase membranes characterized by SEM-EDXS.

**Table S4.** XRD refinements of CNCO-NSFCO41 and CPCO-PSFCO41 powders calcined at 950 °C.

|  | Fluorite (Fm-3m) | Perovskite (Pbnm) | | | |
|---|---|---|---|---|---|
|  | a/Å | a/Å | b/Å | c/Å | Φ/° |
| CNCO-NSFCO41 | 5.42833 (11) | 5.4691(18) | 5.4712(16) | 7.7371(11) | 2.18 |
| CPCO-PSFCO41 | 5.41589 (9) | 5.45183(13) | 5.45462(15) | 7.71663(18) | 2.99 |

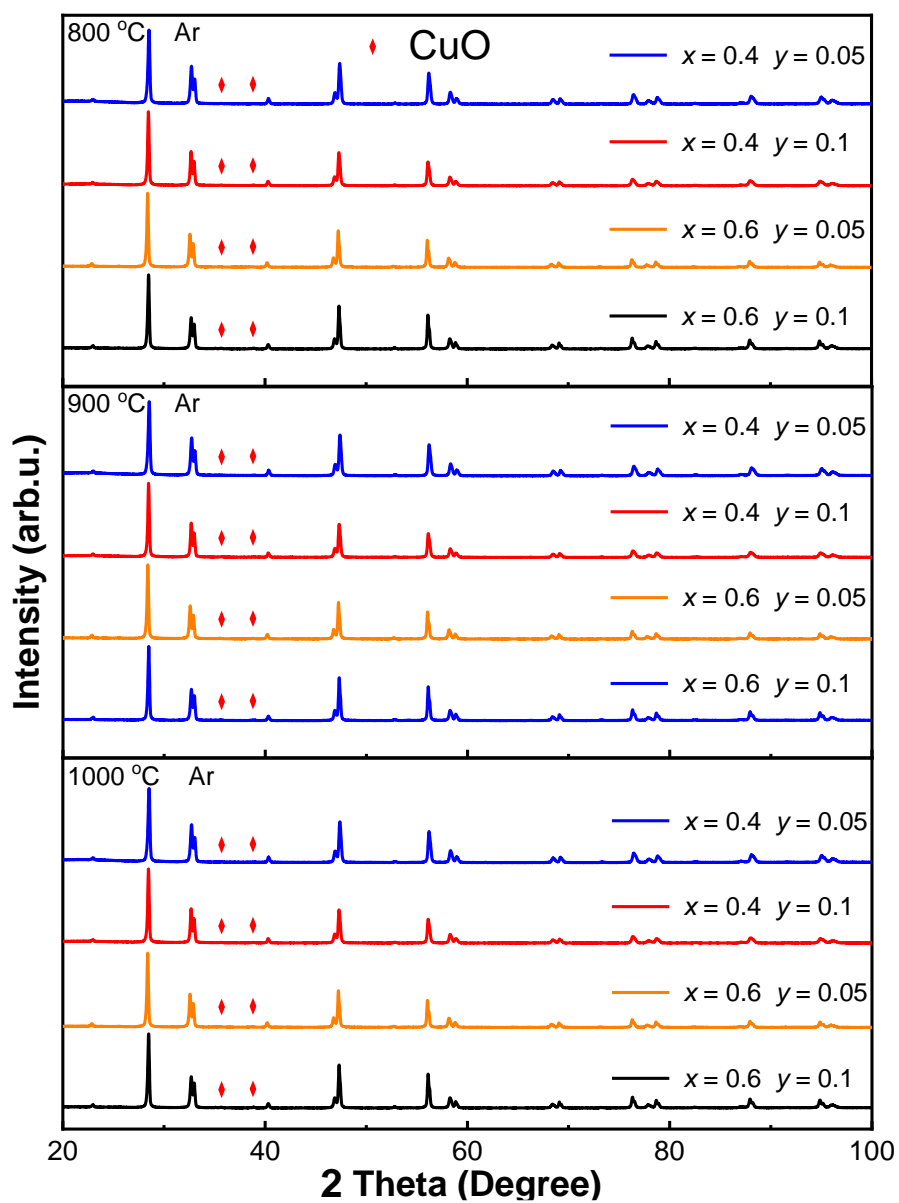

**Figure S6.** XRD patterns of dual-phase powders after 24 hours of treatment in pure Ar at 800 °C, 900 °C, and 1000 °C.

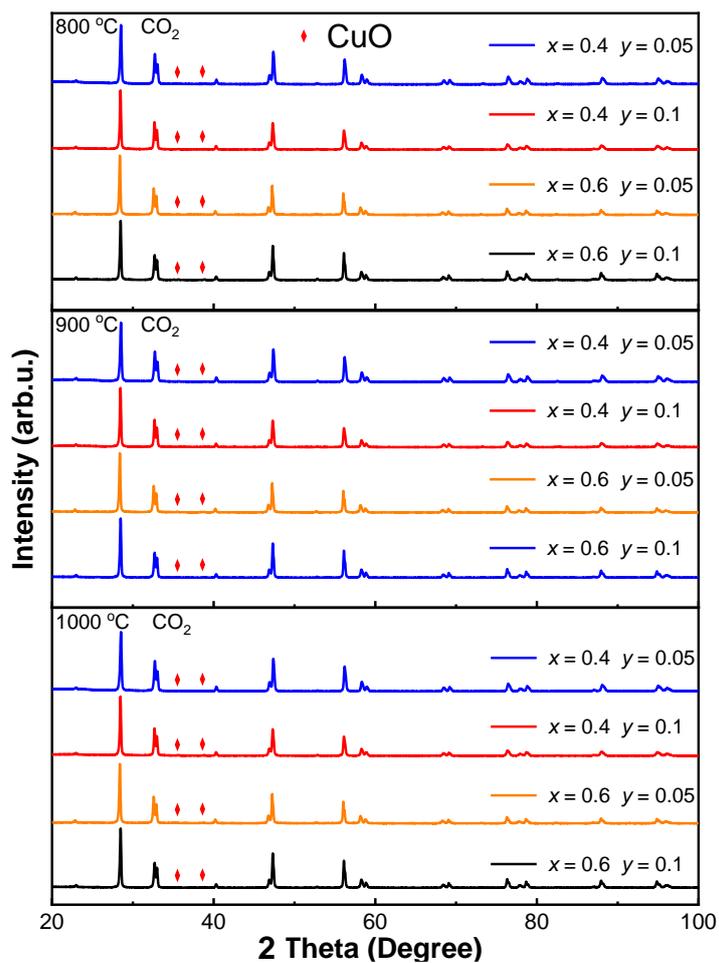

Figure S7. XRD patterns of dual-phase powders after 24 hours of treatment in pure $CO_2$ at 800 °C, 900 °C, and 1000 °C.

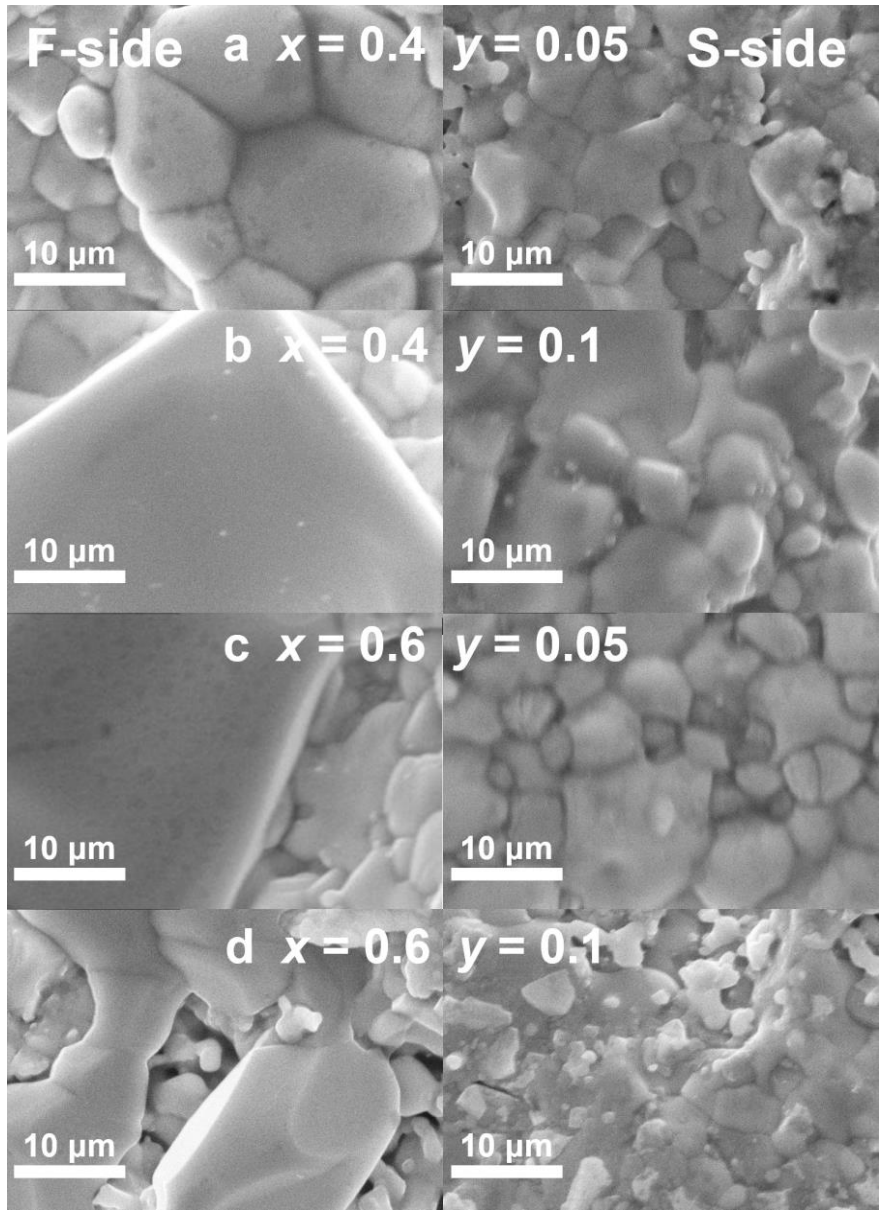

**Figure S8.** Surface topography of both sides of spent dual-phase 60 wt%$Ce_{0.85}Nd_{0.1}Cu_{0.05}O_{2-\delta}$-40 wt%$Nd_xSr_{1-x}Fe_{1-y}Cu_yO_{3-\delta}$ (CNCO-NSFCO; $x$ = 0.4, 0.6; $y$ = 0.05, 0.1) OTMs after the test. The F-side means the feed side, and the S-side represents the sweep side.

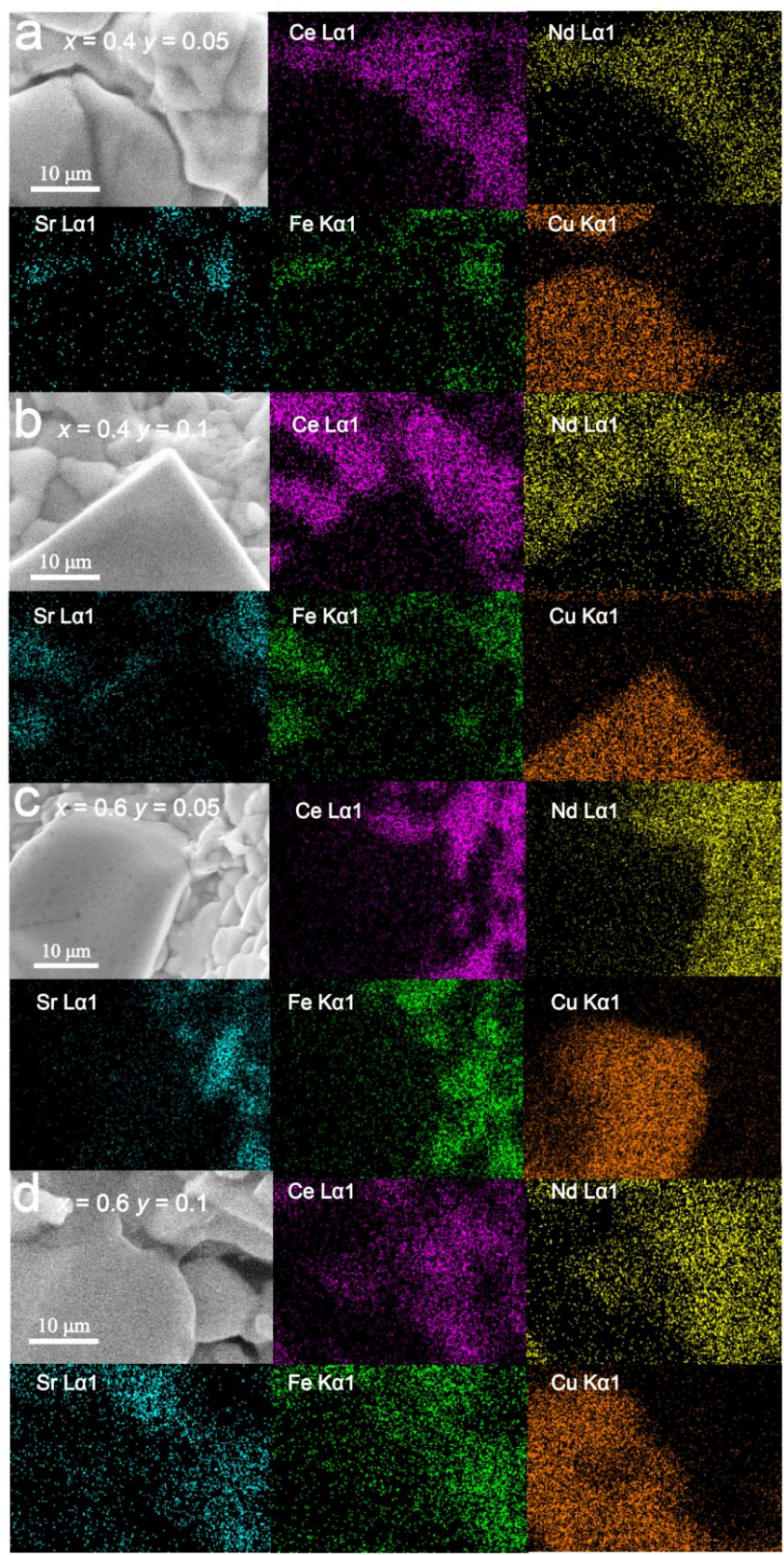

**Figure S9.** Element distribution on the feed sides of spent dual-phase 60 wt%$Ce_{0.85}Nd_{0.1}Cu_{0.05}O_{2-\delta}$-40 wt%$Nd_xSr_{1-x}Fe_{1-y}Cu_yO_{3-\delta}$ (CNCO-NSFCO; $x$ = 0.4, 0.6; $y$ = 0.05, 0.1) OTMs after the test.

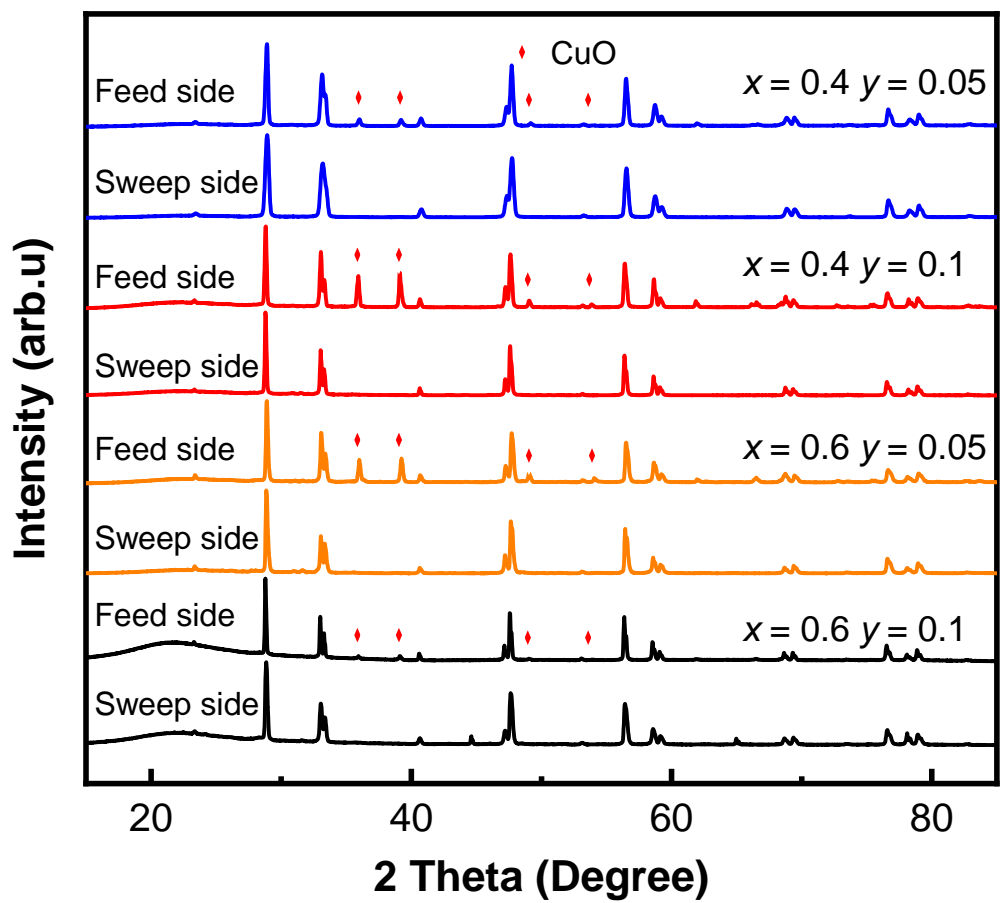

Figure S10. XRD patterns of both sides of spent dual-phase 60 wt%$Ce_{0.85}Nd_{0.1}Cu_{0.05}O_{2-\delta}$-40 wt%$Nd_xSr_{1-x}Fe_{1-y}Cu_yO_{3-\delta}$ (CNCO-NSFCO; $x$ = 0.4, 0.6; $y$ = 0.05, 0.1) OTMs after the test.

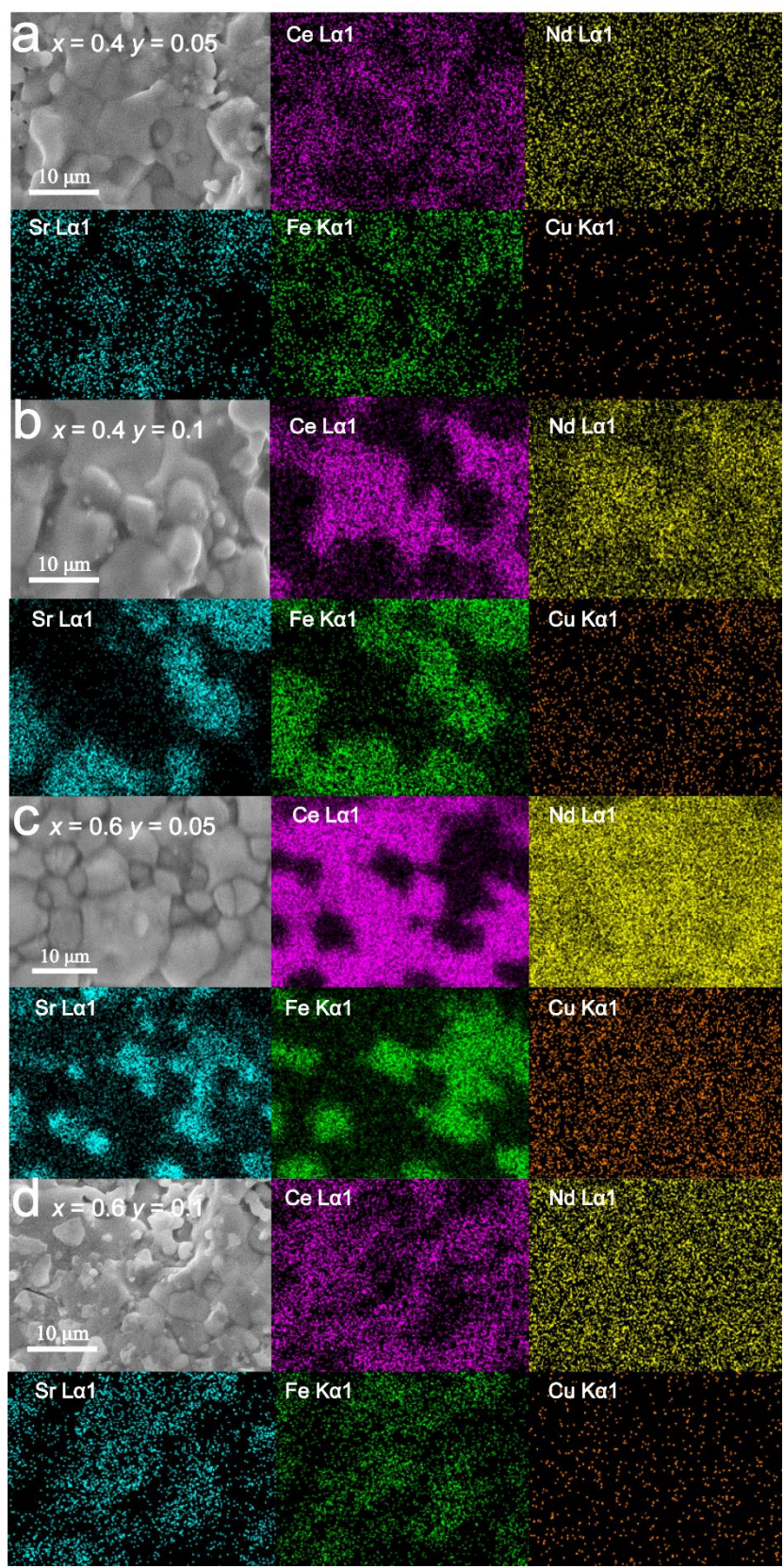

**Figure S11.** Element distribution on the sweep sides of spent dual-phase 60 wt%$Ce_{0.85}Nd_{0.1}Cu_{0.05}O_{2-\delta}$-40 wt%$Nd_xSr_{1-x}Fe_{1-y}Cu_yO_{3-\delta}$ (CNCO-NSFCO; $x$ = 0.4, 0.6; $y$ = 0.05, 0.1) OTMs after the test.

**Table S5. Experimental and theoretical values of the atomic ratios of each element on the surface of fresh membranes.**

| Composition | Side | Element (Atomic percentage) | | | | |
| --- | --- | --- | --- | --- | --- | --- |
| | | Ce | Nd | Fe | Sr | Cu |
| $x = 0.4\ y = 0.05$ | Feed side | 20.83% | 5.02% | 5.84% | 3.35% | 64.96% |
| | Sweep side | 39.85% | 15.51% | 25.05% | 18.30% | 1.29% |
| $x = 0.4\ y = 0.1$ | Feed side | 24.65% | 7.58% | 11.06% | 8.64% | 48.08% |
| | Sweep side | 29.16% | 14.95% | 28.30% | 24.22% | 3.37% |
| $x = 0.6\ y = 0.05$ | Feed side | 15.41% | 8.67% | 10.56% | 7.25% | 58.12% |
| | Sweep side | 49.76% | 19.00% | 19.83% | 10.57% | 0.84% |
| $x = 0.6\ y = 0.1$ | Feed side | 15.41% | 9.06% | 10.75% | 6.35% | 58.42% |
| | Sweep side | 39.27% | 19.73% | 23.56% | 15.63% | 1.81% |